# Evolutionary Optimized, Monocrystalline Gold Double Wire Gratings as a SERS Sensing Platform

Amro Sweedan[1,2], Mariela J. Pavan[1], Enno Schatz[3], Henriette Maaß[3], Ashageru Tsega[4], Vered Tzin[4], Katja Höflich[5], Paul Mörk[6], Thorsten Feichtner[6*] and Muhammad Y. Bashouti[1,2*]

*[1] The Ilse-Katz Institute for Nanoscale Science & Technology, Ben-Gurion University of the Negev, POB 653, Beer-Sheba Campus, Building 51, 8410501, Israel*

*[2] Department of Solar Energy and Environmental Physics, Swiss Institute for Dryland Environmental and Energy Research, J. Blaustein Institutes for Desert Research, Ben-Gurion University of the Negev, Midreshset Ben-Gurion, Building 26, 8499000, Israel.*

*[3] NanoStruct GmbH, Friedrich-Bergius-Ring 15, 97076 Würzburg*

*[4] French Associates Institute for Agriculture and Biotechnology of Drylands, Jacob Blaustein Institutes for Desert Research, Ben-Gurion University of the Negev, Sede Boqer Campus, Be'er Sheva 8499000, Israel*

*[5] Joint Lab Photonic Quantum Technologies, Ferdinand-Braun-Institut gGmbH Leibniz-Institut für Höchstfrequenztechnik, Gustav-Kirchhoff-Str. 4, D-12489 Berlin, Germany*

*[6] Nano-Optics and Biophotonics Group, Experimental Physics 5, Institute of Physics, Am Hubland, University of Würzburg, Germany*

## Abstract

Achieving reliable and quantifiable performance in large-area surface-enhanced Raman spectroscopy (SERS) substrates has long been a formidable challenge. It requires substantial signal enhancement while maintaining a reproducible and uniform response. Conventional SERS substrates are typically made of inhomogeneous materials with random resonator geometries and distributions. As a result, they exhibit several or broadened plasmonic resonances, undesired absorptive losses, and inhomogeneous field enhancement. These limitations diminish the signal strength and hamper reproducibility, making it difficult to conduct comparative studies with high sensitivity. In this study, we propose an approach that utilizes monocrystalline gold flakes to fabricate well-defined gratings composed of plasmonic double-wire resonators, which are fabricated through focused ion-beam lithography. The geometry of the double wire grating substrate (DWGS) was evolutionary optimized to achieve efficient enhancement for both excitation and emission processes. The use of monocrystalline material minimizes absorption losses while enhancing the shape fidelity during the nanofabrication process. The DWGS shows notable reproducibility (RSD=6.6%), repeatability (RSD=5.6%), and large-area homogeneity over areas >$10^4$ µm$^2$. Moreover, it provides a SERS enhancement factor of ≈$10^6$ for 4-Aminothiophenol (4-ATP) analyte and detection capability for sub-monolayer coverage. The DWGS demonstrates reusability, as well as long-term stability on the shelf. Experimental validation with various analytes, in different states of matter, including biological macromolecules, confirms the sensitive and reproducible nature of DWGSs, thereby establishing them as a promising SERS substrate design for future sensing applications.

Sensing with scattered light is invaluable in nearly every aspect of modern life. Raman scattering (RS), in particular,[1] has emerged as a tool for detecting chemicals and identifying them by their fingerprint-like scattering, harnessing the very specific molecular vibrations of the studied molecule. However, unaided conventional Raman spectroscopy suffers from low efficiency ($\eta \approx 10^{-6}$),[2] restricting its application to concentrations greater than 0.01M.

Plasmonic resonances of metallic nanostructures have been extensively studied and optimized since the discovery of surface-enhanced Raman scattering (SERS).[3, 4] It takes advantage of the concentrated light at the surface of metallic nano structures together with the non-linear nature of RS, leading to a signal enhancement up to $10^8$ or even higher.[5] Nowadays, chemical detection,[6] medical[7] and agricultural[8] diagnostics as well as public safety[9] are just a few examples of the ubiquitous fields of application where SERS is already established.

SERS substrates rely on three essential factors to achieve significant signals together with reliable quantitative, and repeatable application: (i) Small gaps within resonant plasmonic structures to attain the highest possible electromagnetic field enhancement; (ii) Large areas covered with dense and regular arrays of resonators to use every incident photon and capture every molecule in low concentrations; (iii) Single nanometer-level fabrication precision to accurately assess the amount of analyte and ensure reproducibility. Consequently, grating, or other long-range resonances can thus be effectively harnessed. Various approaches have been explored, including bottom-up synthesis of agglomerated resonators, (e.g. Refs.[10-12] and many mores in Ref.[4]) top-down fabrication from evaporated gold films,[13-16] or a combination of both approaches.[17-19] While these approaches have resulted in effective SERS substrates, they have not fully met all three crucial aspects (i) – (iii). In addition, we want to address a fourth point that is often overlooked: (iv) The substrate should exhibit plasmonic near-field enhancement at both the excitation wavelength and the emission wavelength of the analyte,[20-22] which can be several tens of nanometers separated from each other.[23, 24]

In this work, we have realized SERS substrates fulfilling requirements (i) – (iv) by applying a novel combination of three main ideas. Firstly, we start with wet-chemically grown monocrystalline gold flakes,[25] a material without lateral grain boundaries, even across large spans of hundreds of micrometers. This enables the highest geometrical control using top-down focused ion-beam (FIB) fabrication and minimizes ohmic losses of the plasmon resonance currents. Secondly, we employ gratings consisting of two wire plasmonic resonators. While large-area grating resonances gather light from expansive regions and provide a first resonance for in-plane scattering,[26] the gaps of the double gratings generate sub-diffraction-limit electrical near-field hotspots at a second resonance, which strongly enhances excitation and re-emission of the Raman-active probe molecules. Finally, we numerically optimize the geometry of this innovative double wire grating substrate (DWGS) using particle swarm optimization (PSO). The optimization process is based on a fitness function that ensures maximal SERS sensitivity for two given wavelengths.

We numerically and experimentally evaluated not only the DWGS concept, but also a single wire grating substrate (SWGS) as well as planar non-structured gold surfaces. For the experiments, all three substrates were realized using both monocrystalline and polycrystalline materials. Our experimental results reveal that the monocrystalline DWGS outperforms all other tested SERS substrates in performance, consistently exhibits reproducible large-area enhancement factors up to several $10^6$, even with a relatively large gap of 15 nm. This advancement overcomes several technological challenges associated with large-scale applications, including reproducibility, repeatability, large area homogeneity, and reusability, among others. Finally, we successfully tested the DWGS on eight different use cases, spanning from chemisorbed, physiosorbed and even gaseous species to proteins and DNA strains. These findings establish SERS based on DWGSs as an auspicious Raman spectroscopy methodology, which can be upscaled by nano-imprint lithography and extended to new applications by utilizing the grating as electrodes in electrochemical environments and sensing technology.

# Results and discussion

**Theory and optimization**

Stokes Raman scattering (RS) involves the excitation of vibrational modes in an analyte by a photon with much higher energy than the vibrational levels. In this process, the photon can be re-emitted with slightly lower energy, as sketched in Fig. 1A after exciting an electron to a virtual energy level. As RS is inelastic, it is highly inefficient,[27] occurring alongside the elastic Rayleigh scattering. For the RS process, both excitation and emission rates are proportional to the field intensity at the analyte position, defined as the local density of states (LDOS).[28] Increasing the LDOS allows to enhance RS, and plasmonic resonances provided by metallic nanostructures do this by concentrating light into sub-diffraction-limited volumes close to their surface. The resulting electromagnetic near-field intensities can surpass diffraction-limited focal fields by several orders of magnitude. This fundamental working principle forms the basis of SERS.

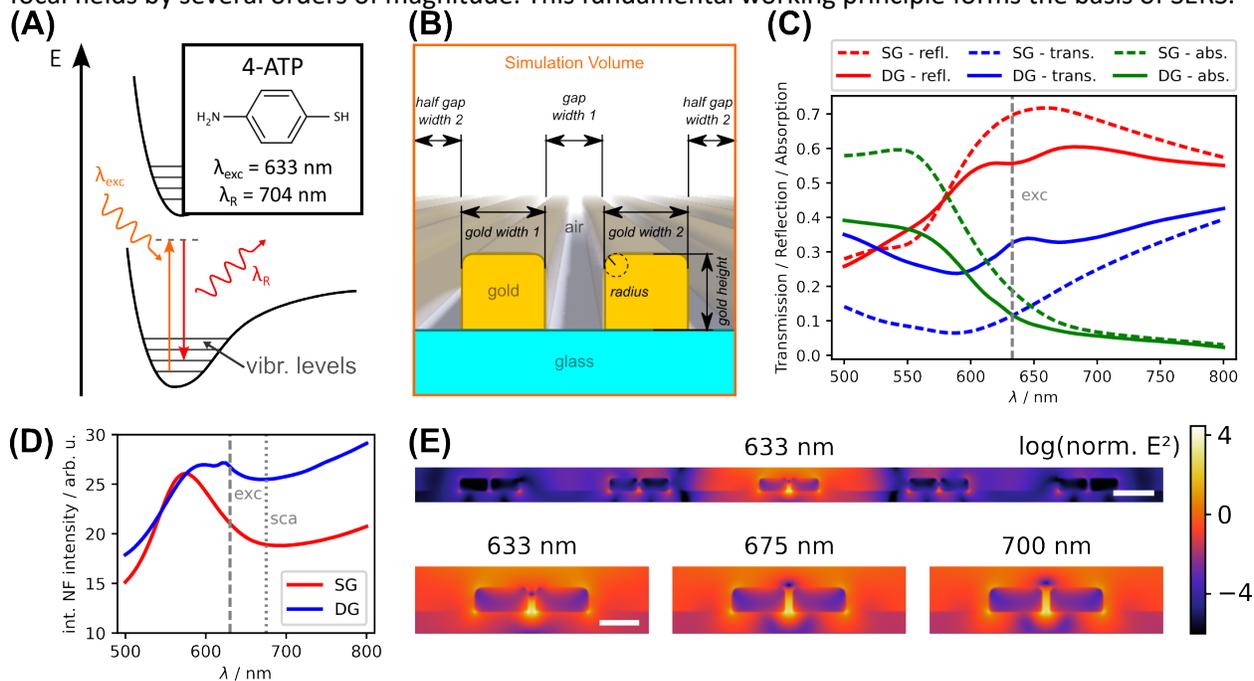

**Fig. 1: Numerical optimization.** (A) simplified energy level scheme of the Stokes-Raman-scattering process. The inset shows the 4-Aminothiophenol (4-ATP) molecule structure, and two possible wavelengths involved in measuring the 1570 cm$^{-1}$ Raman peak. (B) The elementary cell cross-section illustrates the potential possible free parameters for the optimization of DWGS. (C) Spectra of transmission (blue), reflection (red) and absorption (green) for optimized SWGS (dashed lines) and DWGS (solid lines). (D) Integrated near-field intensity enhancement for the optimized SWGS and DWGS. (E) The optimized DWGS SERS substrates near fields are showcased in a vertical cross-section, illuminated from above, at three different wavelengths. It is clearly visible that two different resonances constructively interfere, resulting in high surface field intensities, particularly between and on top of the Au features. Conversely, destructive interference leads to low fields in the central top area of the narrow gap. Scalebar (E): top – 200 nm; bottom – 100 nm.

Most SERS substrates utilizing metal plasmons typically display one or more broad resonances that span both the excitation wavelength $\lambda_i$ and the scattered wavelength $\lambda_s$ as both involved wavelengths only differ in the small energy amount needed to excite a molecular vibration. This allows to express the SERS efficiency in first order approximation to be proportional to $E^4(\lambda_i)$, where $E$ represents the electrical field strength at the location of the molecule to be detected.[4] On the contrary, a well-designed single particle plasmonic resonance can exhibit a typical Q-factor of $Q = \frac{\omega_r}{\Delta\omega} = 15$, where $\omega_r$ is the resonance frequency and Δω is the resonance FWHM. For example, when exciting a 4-Aminothiophenol (4-ATP) molecule[29] with an excitation wavelength $\lambda_i = 633$ nm, one important Raman peak occurs at 1570 cm$^{-1}$, corresponding to an emission wavelength of $\lambda_s \approx 704$ nm (see Fig. 1A inset). A plasmonic resonator optimized for excitation with $Q = 15$ exhibits half its intensity at a wavelength of 654 nm. Consequently, the emission enhancement of such a good plasmonic resonator will be only weak. For this reason, in this study we establish a second resonance in the substrate design to enhance both emission and excitation by using a grating of plasmonic resonators.[26] This is in contrast to work by the Crozier group that introduced the second resonance via a second particle type[20] or by coupling to a second, thin metal layer.[21] We introduce the double wire grating substrate (DWGS) with the cross-section of its elementary cell sketched in Fig. 1B. This choice has several practical benefits: (i) Fabrication via focused ion beam milling is straightforward and even nano-imprint lithography is possible for large-scale high-throughput substrate preparation; (ii) The geometry makes use of small gaps in one dimension to enhance the optical near-fields but still is extended to the second dimension, which makes the substrate polarization sensitive and also will allow the application of an external electrical potential for future electrically modulated SERS (E-SERS) experiments; (iii) Since we only need to simulate in two dimensions, fast evolutionary optimization is possible. The optimization was performed using particle swarm optimization (PSO), a variation of evolutionary optimization implemented in Lumerical/Ansys FDTD Solutions. The fitness function $F$ – needed to compare different DWGS geometries during the evolution – is designed to capture the physics of SERS: enhance the plasmonic near-fields for both excitation $E^2(\lambda_i)$ and emission $E^2(\lambda_s)$ in a volume $V_S$. This volume was set to a layer only a few nanometers thick above the gold surface, as the target analyte is typically situated near the substrate surface through physical or chemical absorption. This leads to:

$$F = \int_{V_s} E^2(\lambda_i) \cdot E^2(\lambda_s) \, \mathrm{d}^3 r \qquad (1)$$

Several parameters have been fixed to factor in fabrication constraints. Gaps are not allowed to be smaller than 15 nm, which represents the smallest reliably realizable gap by means of Ga-FIB milling. The upper grating edge also had a radius of 15 nm due to the point spread function of the Ga-beam (referred to as "radius" in Fig. 1B). Additionally, the height of the DWGS has been fixed at 60 nm, a commonly observed height in our wet-chemically grown gold flakes,[30] which facilitates quick experimental realization after optimization. The light source was modeled as broad-band Gaussian focus from air with a NA = 0.6 to mimic the conditions at the experimental setup. Further details regarding the optimization are briefly commented on in the SI (Fig. S1A, B). The specific results we chose to investigate further are characterized by 'gold width 1' = 139 nm, 'gold width 2' = 136 nm, a small 'gap width 1' = 15 nm and 'gap width 2' = 441 nm. Furthermore, optimization was also conducted on a grating with a SWGS in the elementary cell, resulting in a gold width of 80 nm and a gap width of 20 nm. Fig. 1C-E present the simulated DWGS properties under broad-band optical excitation. Fig. 1C shows far-field spectra, where reflection is proportional to the received signal when detection occurs in the excitation direction due to reciprocity.[31] The DWGS exhibits a transmission up to ~40% due to its large gaps, which are necessary to hit the grating resonance for the given numerical aperture. This additional resonance is visible by an additional feature in both reflection and transmission at the excitation wavelength when compared to the SWGS. This is even more pronounced in the near-field enhancement depicted in Fig. 1D. The near-field spectra are obtained from the region surrounding the central five two-wire building blocks, as depicted in the top field map of Fig. 1E and are normalized to the lateral dimension. The DWGS outperforms the SWGS in terms of both excitation and emission wavelength (denoted by the grey dashed lines in Fig. 1D). Since the excitation near-field peak corresponds to a dip in transmission and reflection, the additional mode is dark and not effectively coupling to the far-field. Instead, it is only coupled via the bright plasmonic double wire mode, scattering into the grating plane. This dark mode has reduced radiative losses and is comparably narrow. The close-up views of the near field in Fig. 1E demonstrate that this additional resonance causes the near fields to cancel out in specific regions within the gap but results in a substantial enhancement closely above the gold surface, as intended by the specific choice of the fitness function $F$. The presence of the initial broad peak in the near fields around 580 nm can be attributed to the hidden constraint of the illumination numerical aperture, as briefly discussed in the SI (Fig. S1C). A closer inspection of the resonances with parameter sweeps and eigenmode analysis will be part of a future publication; however, some aspects can already be found in the SI (Fig. S1D, E). For this work, our focus lies from here on in the experimental aspects of DWGSs.

**Synthesis, fabrication, and characterization**

Both DWGS and SWGS platforms were fabricated via Ga-FIB milling using monocrystalline gold flakes[25] as well as evaporated poly-crystalline gold. Fig. 2A, D illustrate the resulting SWGS and DWGS, along with their respective geometric parameters and illumination conditions. Both grating designs are situated on a glass substrate and are illuminated from the top with a linearly polarized Gaussian beam (spot size $\approx 1.5$ µm). Fig. 2B, C, E and F show scanning electron micrographs (SEM) of the experimentally realized geometries after FIB milling, showcasing high-precision fabrication and high fidelity achieved by using monocrystalline gold.[32] Due to the resolution of the Ga ion beam (~ 5 nm) and sample drift during the large area patterning the intended geometries could only be fabricated with an accuracy of $\pm 5$ nm. Since the plasmonic resonances are comparably broad, this limitation does not hamper the DWGS performance. The bright contrast spots within the gap regions indicate the presence of residual gold particles. Such residues are caused by intermixing during the local physical sputtering in FIB milling or redeposition. Since such particles exhibit an absorption-dominated and polarization-independent resonance slightly above a wavelength of 500 nm, they are not excited by a 633 nm laser as used in the experiments.[33] As a result, their influence can be safely neglected. The inset in Fig. 2E also depicts an electron beam backscatter diffraction (EBSD) map confirming that monocrystallinity is maintained after FIB processing.[32] Since the spatial resolution of EBSD under normal conditions is 20-50 nm, the small 15 nm gap is not visible in the map. FIB nanofabrication results and EBSD of polycrystalline gold results were included in the SI Fig. S2C-F.

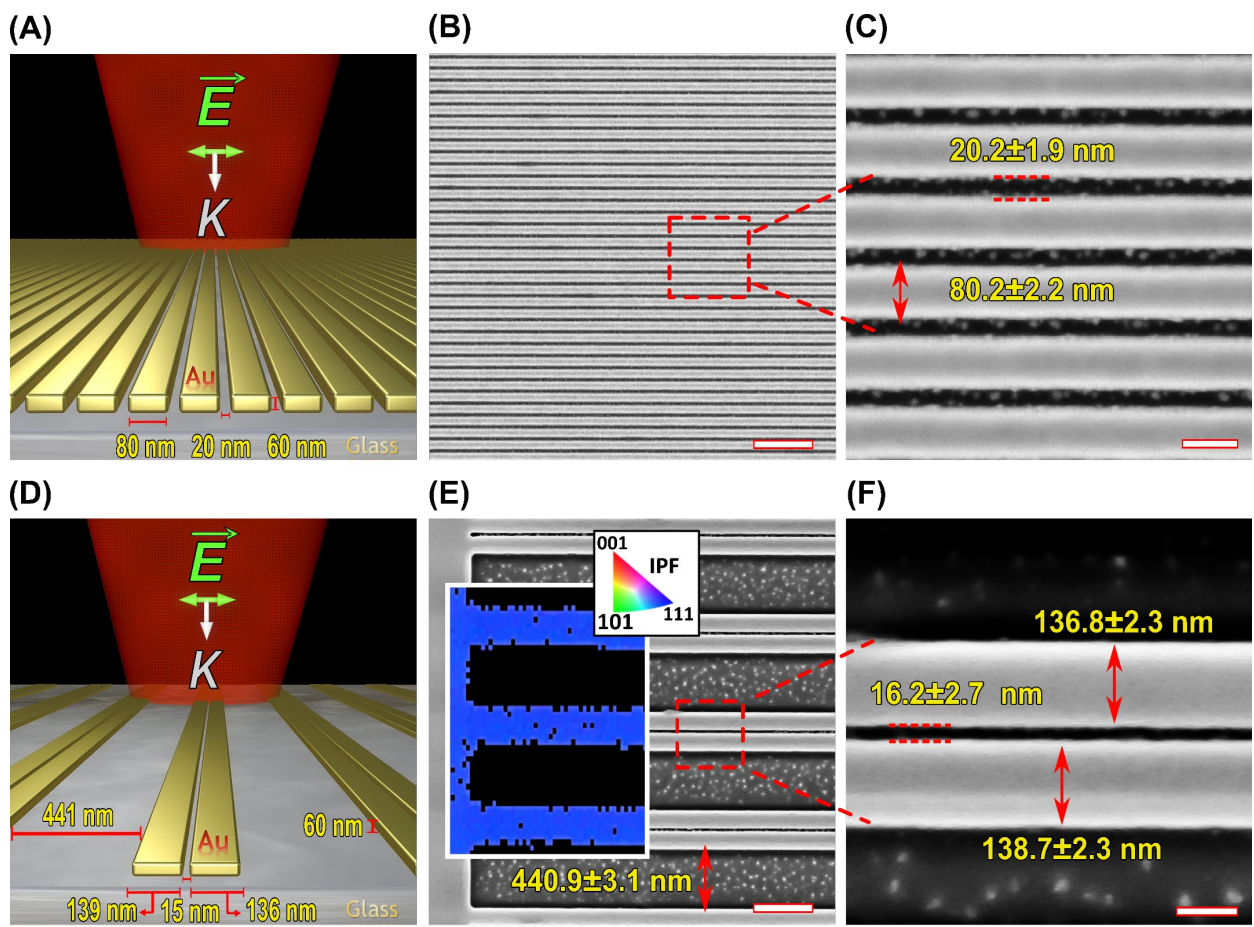

**Fig. 2 Geometry of the SERS platforms.** (A) and (D), 3D schematics of the SWGS and DWGS SERS sensing platforms. The laser polarization and k-vector are represented by green and white arrows, respectively. SEM images of the monocrystalline gold gratings are displayed in (B), (C), (E), and (F) along with the measured dimensions. (B) depicts the SWGS structure, while a higher magnification view is provided in (C). Similarly, (E) shows the DWGS structure, and a higher magnification view is presented in (F). To prove the mono-crystallinity of the final gratings, the inverse pole figure (IPF) map obtained from electron beam scattering (EBSD) measurements is superimposed on the corresponding SEM image in (E). The stereographic triangle of the IPF color map is located on the upper side of (E). Scale bars: (B), (E) – 500 nm, (C) and (F) – 100 nm.

**Spectroscopic properties**

To investigate the SERS-enhancing capabilities of DWGSs, we used several model molecules: 4-Aminothiophenol (4-ATP),[29] Methylene blue (M-blue),[34] Rhodamine 6G (R6G),[35] 3-Mercaptopropionic acid (3-MPA),[36] Methyl benzoate (M-Benzoate),[37] Benzenethiol (BT),[38] as well as biological molecules, including DNA[39] and Bovine Serum albumin protein (BSA).[40]

First, we investigated the SERS signature of 4-ATP self-assembled monolayers (SAM) on DWGS. Our observations in Fig. 3A align consistently with the literature.[41] The two most prominent peaks at 1077 cm$^{-1}$ and 1570 cm$^{-1}$ correspond to the stretching modes of C-S and C=C ring, respectively.[42] These peaks were selected for multiple experiments carried out during the course of this study. To assess the enhancement factor (EF),[5] a reference Raman measurement of 4-ATP was conducted (SI). The resulting SERS analytical enhancing factor for the DWGS falls within the range of ~ $10^6$, as determined by our protocols and sample preparation methods (see experimental section and SI).

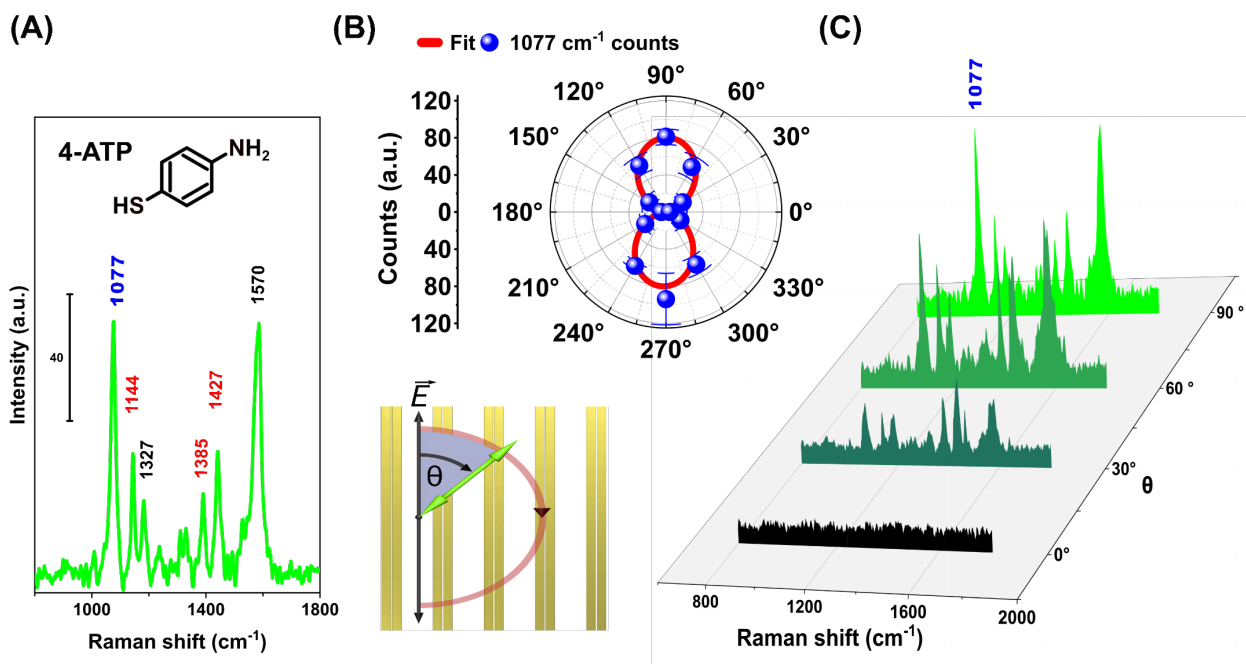

**Fig. 3: Physical properties of DWGS.** In (A), the 4-ATP SERS signature is displayed. In (B), angle-resolved polarized Raman spectra of 4-ATP on DWGS presented as a polar plot, demonstrating the sinusoidal behavior of 1077 cm$^{-1}$ Raman peak intensity with respect to the sample rotation angle θ. The blue dots represent the experimental data, and the red lines depict the theoretical fitting curve. The lower inset in (B) visually represents the changing angle between the electric field and the DWGS gold wires. The angle is obtained by rotating the sample from 0° to 360° while maintaining the laser polarization fixed in Y-axis. Panel (C) provides a plot illustrating the SERS spectrum signature of 4-ATP at different representative angles. Each spectrum represents an average of 3-6 measurements.

We investigated the polarization dependency of the SERS signal from a 4-ATP-DWGS sample (for details see methods section). Fig. 3B and C illustrate the results dependent of $\theta$, the angle between the wires and the excitation polarization. As we excite the gap plasmons in the grating via a dipolar resonance perpendicular to the wires, we expect the Raman signal strength to follow a $\sin^2 \theta$ behavior which is confirmed by the experimental data. Fig. 3B, shows the angle-dependent intensity of the 1077 cm$^{-1}$ peak. The intensity of the peak significantly increased when the polarization angle was varied from 0° to 90°. This confirms the strong enhancement of the Raman signal by the DWGS, which behaves similarly to a flat gold film when the plasmonic gap mode is not excited. To rule out polarization sensitivity of the experiment setup, angle-dependent conventional RS was conducted on bulk 4-ATP analyte deposited on flat gold substrate. The results, along with a polar plot, are concisely discussed in the SI (Fig. S4). As result, a polarization angle of $\theta = 90°$ was selected for all subsequent experiments. Finally, the ability to switch the near-field enhancement by controlling the polarization angle is advantageous in the case of DWGS as it allows for confirming the DWGS plasmonic signal enhancement. When the plasmonic gap mode is not excited, the DWGS behaves similarly to a non-plasmonic flat gold.

The polarization measurement delivered an additional result: During rotation, an ongoing increase in the peaks at 1144 cm$^{-1}$, 1385 cm$^{-1}$ and 1427 cm$^{-1}$ (highlighted in red in Fig. 3A) is observed, which can be seen in Fig. 3C. This increase is attributed to the formation of 4,4'-dimercaptoazobenzene (DMAB) dimers, which is expected to occur due to continuous irradiation and prolonged exposure to oxygen. These dimers are believed to result from oxidation of the 4-ATP molecules induced by the Raman laser. Consequently, this process introduces additional vibrational modes.[41] For the present work we do not investigate this effect further.

To investigate the relationship between morphology, plasmonic properties, and the spatial distribution of gold nanostructures in term of hot-spots, we performed a scanning confocal 2D Raman mapping of 4-ATP SAM on a 5 × 5 µm$^2$ DWGS sample (SI). The mapping reveals the expected hot-spots along the gold wires, and the results are depicted in SI (Fig. S2A-C).

Next, both M-Blue and 4-ATP molecules were investigated using chemical or physical methods respectively, as described in the experimental methods section. These molecules were deposited on six different substrates: flat gold substrates, SWGS and DWGS, each fabricated using both monocrystalline (Fig. 2) as well as polycrystalline gold materials (SI - Fig. S3). The outcomes of these experiments are summarized in Fig. 4.

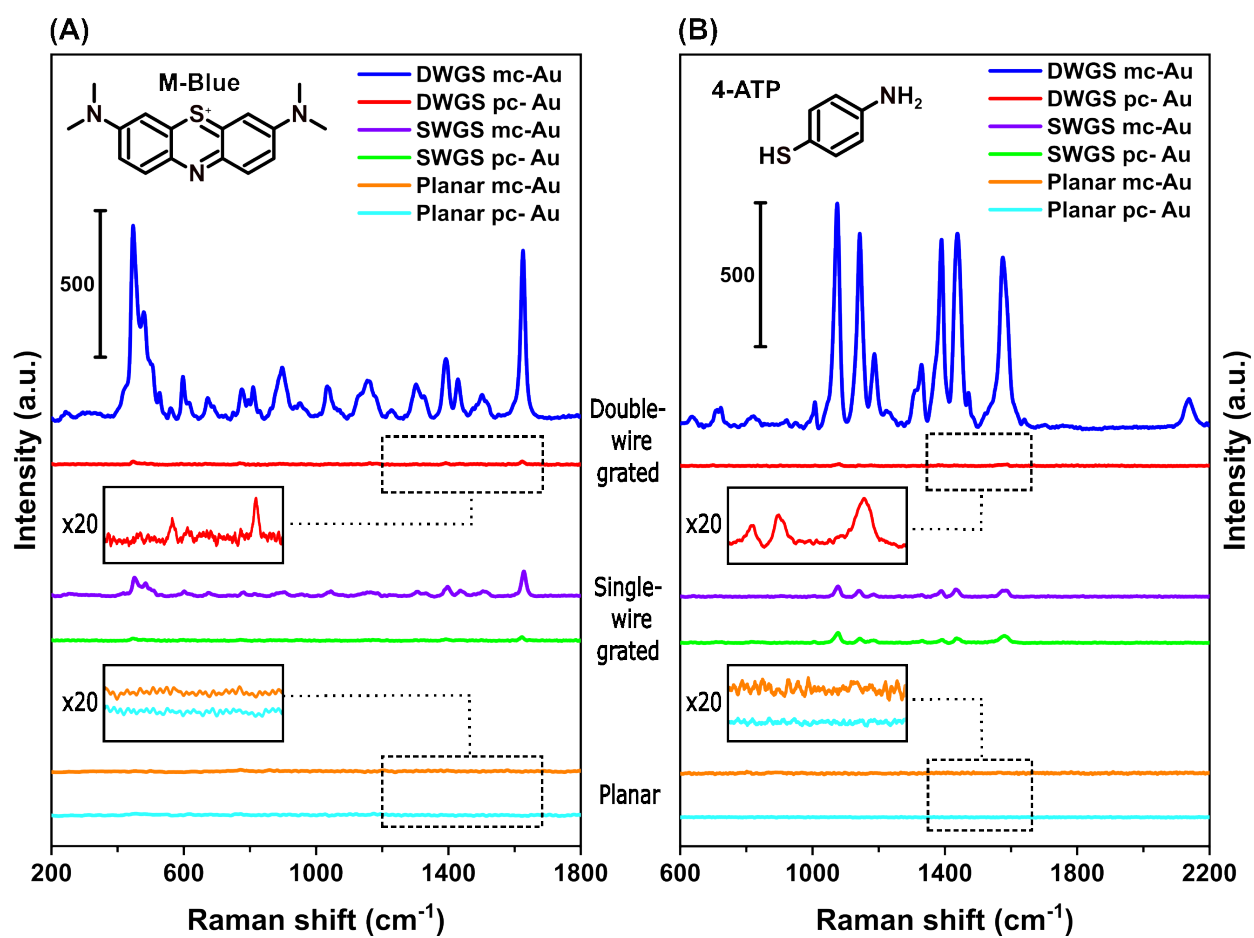

**Fig. 4: Substrate and crystallinity comparison.** Panel (A) displays SERS measurements of M-Blue (1 µM) on dip-casted planar Au (non-grated), single wire grating substrate (SWGS), and double-wire grating substrate (DWGS). Both monocrystalline (mc-Au) and polycrystalline (pc-Au) gold materials were utilized for each substrate type. A zoomed-in view of the interval between 1200 cm$^{-1}$ -1700 cm$^{-1}$ is presented to highlight the relevant peaks. (B) Depicts SERS signals of 4-ATP SAM on the same substrates as in panel (A). Here, a zoomed-in view of the intervals between 1350 cm$^{-1}$ -1650 cm$^{-1}$ is provided.

The DWGS made from monocrystalline gold exhibits the highest SERS signal enhancements for both M-blue and 4-ATP, as shown in Fig. 4A and B. These SERS spectral signatures align with previous studies on these molecules.[41, 43] In contrast, the signals obtained from monocrystalline SWGS substrates are approximately two orders of magnitude lower for both probe molecules. Furthermore, the DWGS monocrystalline substrates demonstrate the ability to detect concentrations as low as ~$10^{-12}$M, as evidenced by a 4-ATP measurement (SI Fig. S5). No SERS signatures were observed from planar monocrystalline or planar polycrystalline gold substrates. Both monocrystalline DWGS and SWGS exhibit higher peak signals when compared to their polycrystalline counterparts. This indicates that the additional surface roughness present on polycrystalline substrates does not sufficiently improve the Raman signal enough to counteract the increased losses caused by the amplified scattering of plasmonic mode currents at grain boundaries.

**Technological properties**

A high-quality SERS substrate should possess several important characteristics to meet the demand of SERS sensing-based technology, as outlined in Refs.[44-46] These characteristics include exceptional sensitivity, homogeneity over a large area, reproducibility in fabrication and signal, long-term stability, cleanliness, and ideally, the ability to be reused. Furthermore, it should be cost-effective to produce. In Fig. 5, we show successful tests of the DWGS for these characteristics, except for the criterion of cost-effectiveness, which remains a challenge for the nano-engineers. Fig. 5A provides evidence of the consistency and repeatability of multiple measurements conducted using the same SERS platform sample under identical conditions. The platform exhibits high repeatability between measurements (RSD = 5.6%) when analyzing the intensities of the 4-ATP molecules at the 1077 cm$^{-1}$ peak using 10 different measurement spots covering about 15% of single substrate area. Furthermore, Fig. 5B attests high reproducibility achieved using 3 different substrates (RSD = 6.6%), for the same peak. All substrate samples were prepared using the same protocol and experimental settings. Fig. 5C supports the claim of long-term signal stability by demonstrating the signal's consistency both after deposition and four months later. Remarkably, the sample was kept at ambient room temperature without any special handling or sealing. The stability of the signal signature can be attributed to the physical and chemical stability of the crystalline gold and the thiol-gold chemical bond. Additionally, Fig. 5D confirms that unused DWGS SERS substrates do not deteriorate over time, demonstrating a long shelf life of at least one year. Samples prepared with these one-year-old unused substrates display a nearly identical Raman signature of the 4-ATP molecules compared to samples prepared with new unused substrates. The substrates were stored at room temperature under ambient conditions without any special treatment or sealing except being boxed against dust contamination. Finally, the reusability of each substrate is made possible due to the inert nature and chemical stability of gold, as well as the intrinsic rigidity of the substrate. Cleaning procedures using various chemical solvents and physical methods can be utilized for the cleaning process of the double grating substrate,[47] enabling its reuse. Mild cleaning procedures are suitable for achieving reusability of pre-used substrates (refer to the experimental methods section). Proof of concept of cleaned reused substrate's reusability is shown at Fig. 5E.

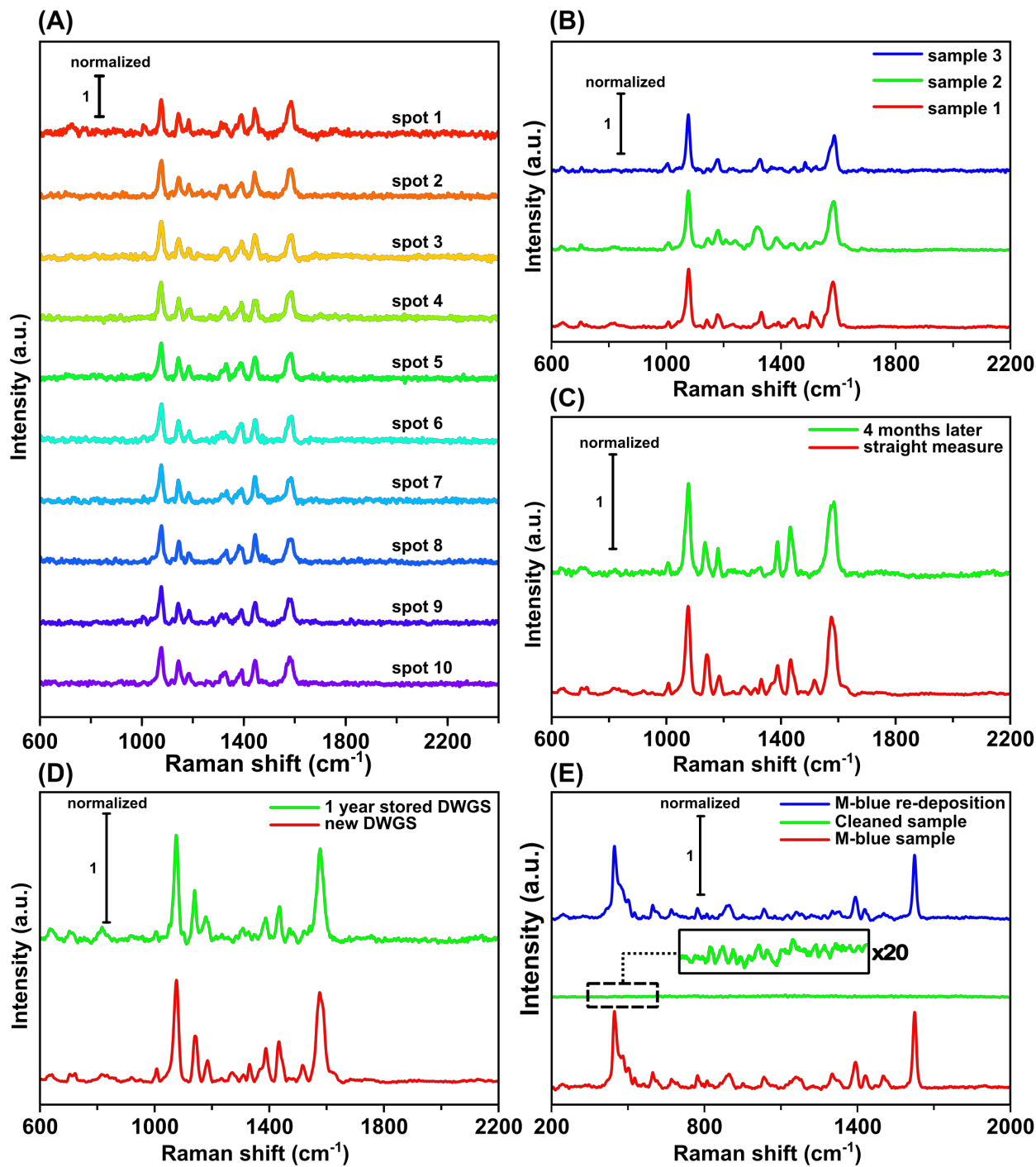

**Fig. 5: DWGS monocrystalline gold platform robustness and essential properties for sensing technology.** (A) SERS signal repeatability on the same SERS platform, showing consistent Raman signals of the 4-ATP SAM for ten sequential measurements on the same sample at different positions. (B) SERS signature reproducibility among three different fabricated DWGS gold substrates samples, all with the 4-ATP SAM. (C) Long-term signal stability of 4-ATP molecule after being left undisturbed at room temperature for 4 months without any special handling. (D) Long shelf time demonstrated with sustained Raman signal generation of the 4-ATP SAM deposited on a one-year-old substrate. (E) DWGS reusability and signal regeneration after a simple cleaning procedure and re-deposition of the M-blue probe molecule. A ×20 zoom-in within the 300 cm$^{-1}$ -600 cm$^{-1}$ interval highlights relevant peaks. Intensities were normalized to improve visual clarity of each experiment.

**Sensing in challenging environments and future possible applications**

Finally, we explore the potential applications of the SERS DWGS in diverse challenging environments and discuss their prospects. These applications encompass the detection of analytes across different states of matter,[48, 49] target molecules in extracts or biological media,[50] and challenging environments such as chemical probes in water,[51] metabolites in solutions,[50] biomedical markers in physiological samples,[52] and volatile gas molecules.[48] We showcase the direct identification of analytes from solution (Fig. 6A), detection of gas molecules through gas flow or solubilization procedures of volatile gaseous molecules (Fig. 6B, C). Furthermore, we investigated the detection of M-Benzoate, a plant metabolic analyte, in plant extracts (Fig. 6D), as well as different biological macromolecules such as DNA (Fig. 6E) and proteins (Fig. 6F).

To detect target molecules in liquids, R6G was solubilized in double-distilled water (DDW) to get 1 µM solution, which was then applied to the DWGS (experimental methods section). Measurements were performed directly in solution by focusing the laser spot through the solution onto the substrate. The presence of molecules not close to the substrate did not interfere with the measurements, as the SERS effect is surface-sensitive within a short range of a few nanometers.[53] We clearly observed the characteristic peaks of R6G, particularly at 612 cm$^{-1}$, 774 cm$^{-1}$, 1310 cm$^{-1}$ and 1361 cm$^{-1}$, which were compared and found to be consistent with the bulk R6G powder material. Furthermore, these peaks were found to be in good agreement with literature.[54]

The detection of volatile gas molecules was demonstrated in two experiments: (1) Direct exposure of DWGS to a carrier gas flow of Benzenethiol (BT) (Argon flow of ~100 ml/min), and (2) solubilization of volatile 3-MPA molecules in a liquid medium in which the DWGS was immersed, thus exposing the substrate to the soluble analyte. The experimental setups for both methods are illustrated schematically in Fig. S6A and B for BT and 3-MPA analytes, respectively, as well as in the experimental section. In the case of solubilizing 3-MPA, volatile molecules were pumped (Argon flow of ~100 ml/min) through a container into a vessel filled with pure DDW containing the SERS substrate. The results are presented in Fig. 6B for BT detection and Fig. 6C for 3-MPA detection. Both 3-MPA and BT exhibited successfully detected peaks that are consistent with literature.[38, 55] Moreover, the SERS peaks-maintained similarity to the Raman peaks of the neat bulk analyte. The SERS signature of the BT analyte showed five prominent peaks at 417 cm$^{-1}$, 998 cm$^{-1}$, 1022 cm$^{-1}$, 1073 cm$^{-1}$ and 1572 cm$^{-1}$, most of which were similar to the conventional Raman spectrum of the bulk BT. Regarding 3-MPA, significant peaks were observed at 295 cm$^{-1}$, 672 cm$^{-1}$, 904 cm$^{-1}$ and 1428 cm$^{-1}$, with the distinctive peaks at 672 cm$^{-1}$ and 904 cm$^{-1}$ that appear both in SERS and neat solution, could be assigned to the CS stretching and C-COO stretching respectively.[55]

The detection of the M-Benzoate molecule in plant extract, as described in the experimental methods section, is summarized in Fig. 6D. The Raman signature of the M-Benzoate molecule was clearly observed in the sample mixture, appearing as a superposition of the pure analyte and the prominent peaks from the plant extract.

The detection of two biological macromolecules exhibited consistency with literature.[56-59] For the DNA SERS signature, the spectrum depends on various parameters such as DNA sequence, orientation, and sample preparation.[59] However, certain peaks, including the one at 670 cm$^{-1}$, are commonly observed due to similar functional groups of DNA building blocks.[58] The DNA SERS signature obtained using DWGS provided several distinct peaks, as shown in Fig. 6(E), including peaks at 670 cm$^{-1}$, 988 cm$^{-1}$, 1029 cm$^{-1}$, 1340 cm$^{-1}$ and 1538 cm$^{-1}$. The major peaks at 670 cm$^{-1}$, 988 cm$^{-1}$, 1029 cm$^{-1}$ could be attributed to guanin base, sugar backbone, and cytosine, respectively.[56, 58] The Raman signals of Bovine Serum Albumin (BSA) were easily observed using SERS, as depicted in Fig. 6F, compared to the bulk signal of BSA powder and previously reported literature.[57] Particularly, distinct peaks at 1000 cm$^{-1}$, 1560 cm$^{-1}$ and 1580 cm$^{-1}$ were assigned to phenylalanine and tryptophane amino acid.

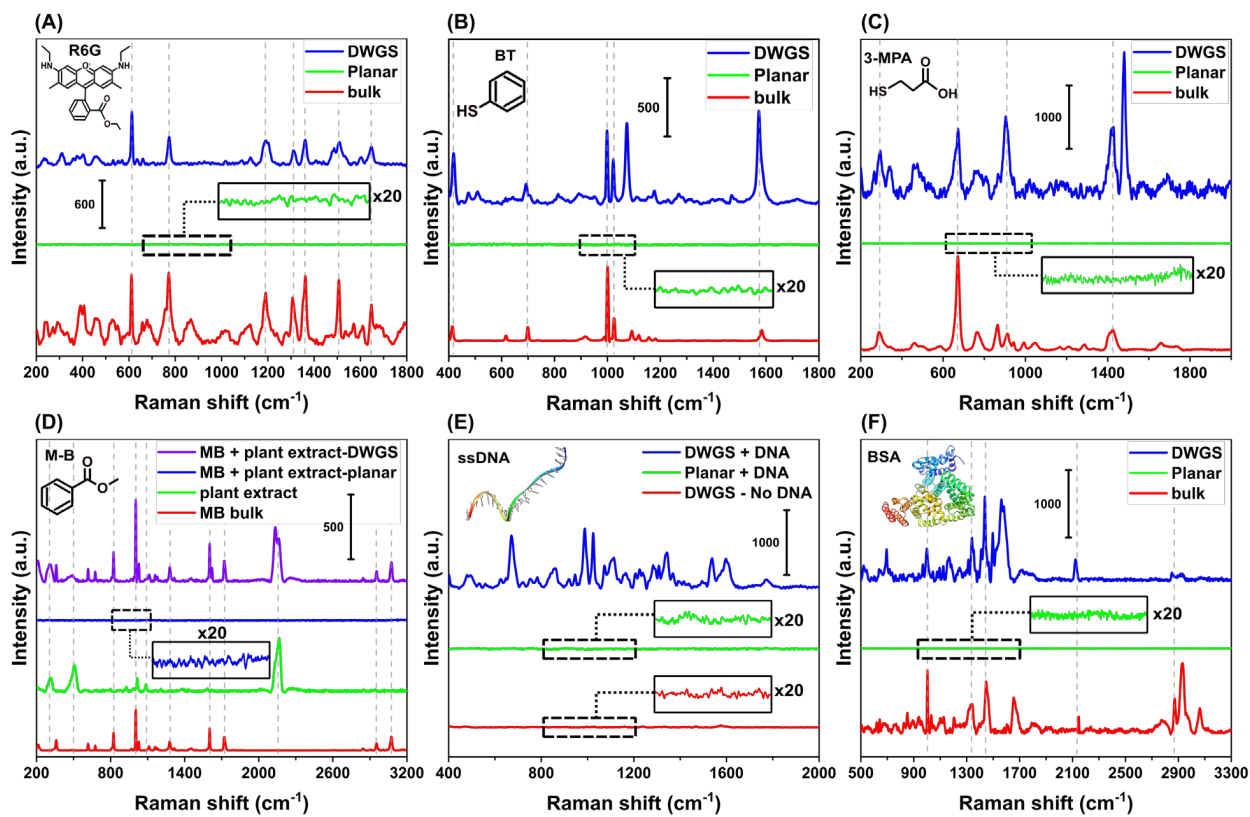

**Fig. 6: Showcases the potential future applications of DWGS platform.** The substrates are: Planar Au flake (non-grated), and double wire grating substrate (DWGS). (A) illustrates the platform's ability to detect R6G from solution. (B) shows the direct detection of the BT probe molecule from the gas phase. (C) presents the detection of volatile 3-MPA analytes that were solubilized in a liquid medium containing the substrate, followed by detection after substrate drying. (D) serves as a proof of concept by detecting the Methyl benzoate volatile molecule in plant extract sample. (E) and (F) depict the detection of SERS signatures of biological macromolecules in solution, specifically ssDNA and BSA protein, respectively. Zoomed-in views with ×20 magnification are provided to emphasize the relevant peaks. The zoom intervals are as follows: (A) 550 cm$^{-1}$ – 850 cm$^{-1}$, (B): 900 cm$^{-1}$ – 1100 cm$^{-1}$ (C) 800 cm$^{-1}$ – 1800 cm$^{-1}$, (D) 800 cm$^{-1}$ – 1100 cm$^{-1}$, (E) 800 cm$^{-1}$ – 1200 cm$^{-1}$, (F) 900 cm$^{-1}$ – 1700 cm$^{-1}$. The model of ssDNA is based on the coordination of DNA polymerase-clamp-exonuclease determined by Cryo-EM,[60] while the BSA model is based on Crystal Structure of Bovine Serum Albumin.[61] Molecular graphics and analyses of biomolecules were performed with UCSF Chimera software.[62]

## Conclusion

In summary, we have introduced the double wire grating substrate (DWGS), an innovative SERS platform based on a grating of two-wire waveguides – double resonant nanostructures fabricated from monocrystalline gold using focused ion beam lithography. Through careful design of the fitness function for evolutionary optimization, these DWGSs achieve large local field enhancement for both excitation and emission processes involved in Raman scattering. The monocrystalline material optimizes near- and far-field properties and enables precision and fidelity in the nanofabrication process over a large area, with enhancement factors reaching several $10^6$ and the ability to detect concentrations as low as ~$10^{-12}$ M in solution. The dipolar double wire resonances allow to "switch" the surface enhancement via the excitation polarization angle. The DWGS platform presents numerous technological advantages, namely high sensitivity, and noteworthy homogeneity over large areas (>$10^4$ µm$^2$), notable repeatability (RSD = 5.6%), and reproducibility (RSD = 6.6%). Moreover, it is long term stable up to at least one year under ambient conditions, requiring no special handling or sealing, and can be reused multiple times without a noticeable loss in performance. To highlight its practical usability, we successfully applied the DWGS platform with both chemical and physical deposition methods, enabling the detection and identification of trace amounts of analytes across different states of matter. This includes gaseous analytes, plant extracts, and biological macromolecules. The investigation of DWGS in this study reveals its potential as an advanced SERS platform, capable of enhancing analytical capabilities in various fields. For even more SERS enhancement, the use of Helium FIB will allow for gaps down to a few nanometers which is expected to localize and enhance the fields even further.

## Experimental methods

### Simulation

Simulations were performed with Ansys Lumerical FDTD Solutions (Release 2021 R2.5) in a purely 2D simulation area. Excitation was performed via a broadband pulse with a 500 – 800 nm wavelength interval and a thin lens approximation of an NA = 0.6 objective to mimic the experimental conditions. The materials employed were glass as user defined dielectric with fixed index at n = 1,518 representing borosilicate glass, and gold as a fit of the Etchegoin material model in its corrected version.[63] The grating contained 21 double-grating elementary cells, which ensures convergence against a hypothetical infinite grating. Only the five central elementary cells were evaluated for the fitness function used in the particle swarm optimization (100 generations containing 20 individuals) of the grating parameters. A mesh refinement area with fixed 2 nm Yee cell size was placed around all gold within the simulation up to 20 nm distance. Distance from resonator to the boundary PML (standard settings) was set to be 500 nm to ensure a distance larger than half of the maximal wavelength.

### Gold synthesis

Gold flakes were synthesized via a modified wet chemical procedure based on Ref.[30] Briefly, in a typical procedure, Hydrogen tetrachloroaurate trihydrate (Aldrich, SKU: 520918) was dissolved in 1 ml DDW (1M final concentration) and mixed by stirring in 20 ml ethylene glycol (Aldrich, SKU: 102466), serving as the solvent in 50 ml polypropylene centrifugation tube. A microscopy coverslip was used as a substrate for the gold flake growth. Prior to immersion in the gold solution, the substrate was thoroughly cleaned in ethanol and acetone, followed by rinsing with DDW and drying with a stream of Nitrogen ($N_2$). Two substrates were immersed into the homogeneous gold solution back-to-back. Subsequently the solution was heated in an oven and kept at 90 C° for 18 hr. Then, the substrates were removed from the solution, rinsed with ethanol and water, and dried with a $N_2$-stream. Typically, the resulting flake thickness ranges between 40 and 80 nm, which can be measured by fitting a broadband transmission spectrum.[30] Gold flakes with a thickness of 60 nm were identified and transferred onto partly evaporated gold / glass substrates (marker structures & ground contact) with a PMMA-mediated method[64] to ensure ground contact for subsequent focused ion-beam milling. Evaporated gold substrate was produced by thermally depositing a gold film on silicon or glass substrates under a high vacuum. A 5 nm Titanium / Chromium adhesive layer is deposited on the substrate, followed by a 60 nm gold layer.

**FIB fabrication**

Focused ion beam milling is conducted using the Gallium beam of a Zeiss ORION NanoFab system, operating at an acceleration voltage of 30 kV and a beam current of 10 pA. For the larger gaps, the beam is scanned with 35 passes, a 1µs dwell time, and a serpentine direction, applying an ion dose of 4.7 mC/cm$^2$. The small gaps were fabricated using a single line pass and a dwell time of 300 µs.

**SERS Model molecule sample preparations**

4-ATP (Aldrich, SKU: 422967) was chemically deposited using the standard gold thiolation method in solution.[65] The substrate was immersed in a glass vial containing 5 ml of 1.5 mM ethanol solution of 4-ATP at room temperature for 24 hours. This incubation led to the formation of a self-assembled monolayer of 4-ATP on the gold substrates. For calculating the enhancing factor, DWGS substrate was immersed in 5 ml solution 4-ATP with concentration of 1.5 mM in Abs. Ethanol for 25 hours. The reference was obtained by drop-casting 0.5 µl of 15 mM 4-ATP (in Abs. Ethanol) on a planar gold surface, followed by solvent drying in a hood for 10-30 min. The 0.5 µl ethanol solvent cover nearly 1mm$^2$ of the gold surface. Both measurements were done under identical experimental conditions and preparation conditions. The enhancing factor was then calculated by averaging the 1077 cm$^{-1}$ integrated peaks from 3 spectra.

M-Blue (Aldrich, SKU: 66720) molecules were physically deposited by dip-casting the sample into corresponding concentrations of the analyte in DDW (Ranging from µM to mM concentration) for 2 hours. Subsequently, the sample was snap-rinsed 10 times in clean Ethanol and 10 time in DDW, then dried with a stream of Nitrogen (N$_2$).

For sample reuse, the surface was cleaned by mild basic piranha treating for 20 minutes at 60 °C using a 4:1:1 mixture of DDW:NH$_4$OH:H$_2$O$_2$. Adjusting treating time or mixture ratio is recommended for desired cleaning without damaging the sample.

Liquid-phase SERS direct measuring of R6G (Aldrich, SKU: R4127) was obtained by preparing a 1 µM solution in DDW, applying 1 µl on the substrate surface, covering it with a microscopy glass coverslip, and measuring directly.

The setup for sensing gaseous volatiles of BT (Benzenethiol; Aldrich, SKU: T32808) molecule is shown schematically in the SI (Fig. S5 A). The gas was actively flown through a tube to the SERS substrate. A 0.5 ml sample of BT (~17 % (v/v) in ethanol) was applied into a warmed container (*3*0 °C). An active flow of Argon (~100 ml/min) was maintained for 30 min, followed by removing the substrate and performing measurements.

For sensing solubilized gaseous molecules, 3-MPA (Aldrich, SKU: M5801) was drop-casted (50µl) and pumped from a warmed container (30 °C) into an adjacent chemical beaker containing the SERS substrate immersed in pure DDW as shown in the SI (Fig. S6B). The inert pumping flow was carried out using active argon flow (~100 ml/min) for 1 hour, followed by drying the sample for 10 minutes in a hood and measuring the 3-MPA attached to the SERS substrate.

Both biological molecules were prepared in a similar way. Both the single-strand DNA (Sequence: 5'CGATCCACCTCCGGAACCTCCACCTTTTTCGAATTGTGGATGACTCCAAGCGGAGCCGCCTTTACCCGGGGACAG GGAGAG -3') and the bovine serum albumin (BSA) were prepared by applying of 1 µl analyte solution (1 µM in DDW) to the sample, covering it with a microscopy glass coverslip and measuring directly.

The plant sample extract was prepared following methods described in Ref. [66]. Briefly, Tomato plants (*Solanum lycopersicum*) of the cultivar named Ikram (Syngenta) were grown for four weeks under controlled growth conditions (16 h/8 h light/dark, 24 ± 2°C). Fresh leaves then harvested, frozen, and homogenized. Metabolites were extracted using an organic phase composed of Methanol, Chloroform and MiliQ $H_2O$ ratio of 2.5 :1 :1 (v/v) respectively, to remove chlorophyll and avoid fluorescence interference. The extract was sonicated for 10 min at room temperature, centrifuged, and the upper supernatant was collected. Next, it was then mixed with 1 mM M-Benzoate (Aldrich, SKU: 8223301000) solution (In Ethanol) in a 1:50 v/v M-Benzoate solution to extract plant extract. The mixture was applied on the SERS substrate, covered with microscope coverslip, and directly measured.

**Raman measurements and data analysis**

Raman spectra were collected using a confocal Horiba LabRam HR Evolution with a 633 nm excitation laser and a power at the sample 0.09 mW (except of 4-ATP $10^{-12}$ M experiment (Si), the power at the sample was 0.5 mW). The measurements were performed using 50x LWD objective (Olympus LMPlanFL-N, NA = 0.5, spot size about 1.5 µm). The measurements were taken with a 600 gr/mm grating and a 100 µm confocal hole. The instrument was equipped with a Syncerity multichannel 1024 x 256-pixels CCD detector cooled to a temperature of -60 °C. The approximate spectral resolution was 1.3 $cm^{-1}$. Each spectrum was acquired for 20 seconds with 3 accumulations. System calibration was performed by assessing the first-order phonon band of Silicon (111) substrate at 520 $cm^{-1}$. Most measurements were conducted using the same experimental settings. In cases where different exposure times or laser intensities were used, the data were normalized to obtain counts per 1 sec and 1 mW, allowing for comparison under adjusted uniform setting. LabSpec software (version 6.5.2.11) was used to operate the spectrometer. To assess reproducibility and repeatability, at least three consecutive accumulation spectra were acquired back-to-back at the same location, and a second set of at least three repeat spectra was collected immediately thereafter at a slightly offset location (1-3 µm). The average SERS intensities and relative standard deviations of the whole spectrum were calculated based on 3-25 SERS spectra, with the relative standard deviation (RSD) calculated for specific peaks. Baseline Raman spectra from the substrates were obtained prior to analyte deposition. Raman data processing, including spectra smoothing, baseline correction, and normalization was performed using OriginPro 2023. The smoothing algorithm was Savitzky-Golay with second polynomial order. Base line subtraction was done using a spline interpolating method with second derivate. After performing a basic processing for each spectrum, such as despiking. Spectra obtained from the same experiment (at least from 3 points) were averaged and plotted as one spectrum. To improve visual clarity, the SERS intensities were normalized using a min-max spectra normalization approach.

To measure the polarization dependency of the 4-ATP-DWGS SERS signal, a continuous 360° manual micrometer-based rotation stage was used. The polarization direction of the excitation laser along the *y*-axis of the sample was maintained throughout the experiment. The DWSG substrate was rotated in steps of 30°. Data were fitted by $y(\theta) = y_0 + A \sin^2\left(\pi \left(\frac{\theta - \theta_0}{w}\right)\right)$. Where $\theta_0$ was fixed to 0°.

**SEM and EBSD characterization**

SEM imaging was conducted using a Thermo Fisher Verios 460L field-emission scanning electron microscope, equipped with an electron backscatter diffraction (EBSD) detector (OXFORD instruments). Images were acquired in the secondary electron mode, with an accelerating voltage of 3 kV, a beam current of 25 pA, and a working distance of 5.9 mm. For EBSD analysis, the sample was tilted by 70°, using an electron beam with an acceleration voltage of 20 kV and a beam current of 0.8 nA. The backscattered electrons, which undergo partial diffraction, enabled the visualization of characteristic Kikuchi patterns. The OXFORD Channel 5 software was employed to analyze these patterns and determine the relevant 3D orientation information for each pixel.

## Data availability

Data available within the article or its supplementary materials. Raw data can be provided upon request – please contact the corresponding authors.

## Author information

**Corresponding Authors**

**Muhammad Y. Bashouti** – *Department of Solar Energy and Environmental Physics, Swiss Institute for Dryland Environmental and Energy Research, J. Blaustein Institutes for Desert Research, Ben-Gurion University of the Negev, Midreshset Ben-Gurion, Building 26, 8499000, Israel. The Ilse -Katz Institute for Nanoscale Science & Technology, Ben-Gurion University of the Negev, POB 653, Beer-Sheba Campus, Building 51, 8410501, Israel. Department of Physics, Ben-Gurion University of the Negev, Beer-Sheva 84105, Israel. Email: bashouti@bgu.ac.il*; ORCID: *https://orcid.org/0000-0002-0371-7088*.

**Thorsten Feichtner** – *Nano-Optics and Biophotonics Group, Experimental Physics 5, Institute of Physics, University of Würzburg, Am Hubland, 97074 Würzburg, Germany. Email: thorsten.feichtner@physik.uni-wuerzburg.de*; ORCID: *https://orcid.org/0000-0002-0605-6481*.

**Authors**

**Amro Sweedan** – *The Ilse -Katz Institute for Nanoscale Science & Technology, Ben-Gurion University of the Negev, POB 653, Beer-Sheba Campus, Building 51, 8410501, Israel. Department of Solar Energy and Environmental Physics, Swiss Institute for Dryland Environmental and Energy Research, J. Blaustein Institutes for Desert Research, Ben-Gurion University of the Negev, Midreshset Ben-Gurion, Building 26, 8499000, Israel. Email: sweedan@post.bgu.ac.il*; ORCID: *https://orcid.org/0000-0001-9623-9726*.

**Mariela J. Pavan** – *The Ilse -Katz Institute for Nanoscale Science & Technology, Ben-Gurion University of the Negev, POB 653, Beer-Sheba Campus, Building 51, 8410501, Israel. Email: marielap@bgu.ac.il*; ORCID: *https://orcid.org/0000-0002-9512-2160*.

**Enno Schatz** – *NanoStruct GmbH, Friedrich-Bergius-Ring 15, 97076 Würzburg. Email: enno.schatz@nanostruct.eu*; ORCID: *http://orcid.org/0000-0001-5241-5003*.

**Henriette Maaß** – *NanoStruct GmbH, Friedrich-Bergius-Ring 15, 97076 Würzburg. Email: henriette.maass@nanostruct.eu*.

**Ashageru Tsega** – *French Associates Institute for Agriculture and Biotechnology of Drylands, Jacob Blaustein Institutes for Desert Research, Ben-Gurion University of the Negev, Sede Boqer Campus, Be'er Sheva 8499000, Israel. Email: ashageru@post.bgu.ac.il.*

**Vered Tzin** – *French Associates Institute for Agriculture and Biotechnology of Drylands, Jacob Blaustein Institutes for Desert Research, Ben-Gurion University of the Negev, Sede Boqer Campus, Be'er Sheva 8499000, Israel. Email: vtzin@bgu.ac.il. https://orcid.org/0000-0002-5912-779X.*

**Katja Höflich** – *Joint Lab Photonic Quantum Technologies, Ferdinand-Braun-Institut gGmbH Leibniz-Institut für Höchstfrequenztechnik, Gustav-Kirchhoff-Str. 4, D-12489 Berlin, Germany. Email: katja.hoeflich@helmholtz-berlin.de. https://orcid.org/0000-0003-4088-2928*

**Paul Mörk** – *Nano-Optics and Biophotonics Group, Experimental Physics 5, Institute of Physics, University of Würzburg, Germany. Email: paul.moerk@stud-mail.uni-wuerzburg.de. https://orcid.org/0009-0005-1003-1446*

**Author contributions**

The manuscript was written through the contributions of all authors. All authors have given approval to the final version of the manuscript. T.F. conceived the presented idea, including the initial substrate geometry, performed numerical design optimization, and numerically analyzed the plasmonic properties of the final DWGS. A.S. and M.Y.B. planned and carried out the SERS experiments, methodology validations, software analysis, wrote the first draft including models, and took the lead in writing the manuscript. M.Y.B. and T.F. co-supervised the project and acquired the funding. M.J.P. assisted in analyzing the results and revising the manuscript. E.S., K.H., and H.M. managed the top-down fabrication, contributed to result analysis, and participated in discussions. V.T. and A.T. performed the plant extract preparation. P.M. revised the manuscript and participated in discussions.

**Note**

The authors declare no conflict of interest.

**Acknowledgments**

We would like to express our sincere gratitude to the WITec company for their collaboration, particularly N. Hafi, for his assistance mainly in conducting the mapping experiments and relevant data analysis. We would also like to extend our thanks to I. Benhar at Tel Aviv University in Israel for (i) his input in designing the gaseous volatiles molecule sensing experiment chamber, (ii) for his team particularly L. Nahary, for support during the biological experiments. We thank the Ilse Katz Institute for Nanoscale Science & Technology center for providing the infrastructure for SEM, EBSD, and Raman characterization. We also express our gratitude to A. Shalabny for his support with the experiments and lab work. A. Sweedan is grateful for the institutional Ph.D. scholarship from Ben-Gurion University of the Negev and from IIse-Katz Institute for Nanoscale Science & Technology.

This research received financial support from various sources, T.F. acknowledges financial support from the European Commission through the Marie Skłodowska-Curie Actions (MSCA) individual fellowship project PoSHGOAT (project-id 837928) and participation in CA19140 (FIT4NANO), supported by COST (European Cooperation in Science and Technology). M.Y.B. acknowledges the Goldinger-Jewish Fund for the Future (no. 87703911) and NIFA-BARD (no. 81122911). V.T. AND M.Y.B. acknowledges the financial support of Israel Ministry of Agriculture (no. 16-38-0029), Biobee Sde Eliyahu Ltd., and ICA.

## Supporting information available

Abbreviations used in the article are summarized in table S1. A real-time digital camera video (Avi) of 4-ATP-DWGS spectrum intensity changes along the grating. Simulation based optimization of double wire geometry, including the development of grating fitness generations, far-field pattern, and grating dimension optimization (Fig. S1). Enhancing factor calculation. Additional experimental details and 2D Raman mapping of 4-ATP-DWGS (Fig.S2). SEM and EBSD results, Au flakes SEM, and evaporated Au film FIB fabrication (Fig. S3). 4-ATP conventional Raman against different polarization angels (Fig. S4). SERS spectrum and calibration curve of M-Blue at different concentrations (Fig. S5), 4-ATP picomolar concentration detection (Fig. S5) and schematics of gaseous molecules sensing experimental setup (Fig. S6). Photoluminescence and background signal of DWGS substrate Fig. S7 and Fig.S8

**For Table of Contents Only**

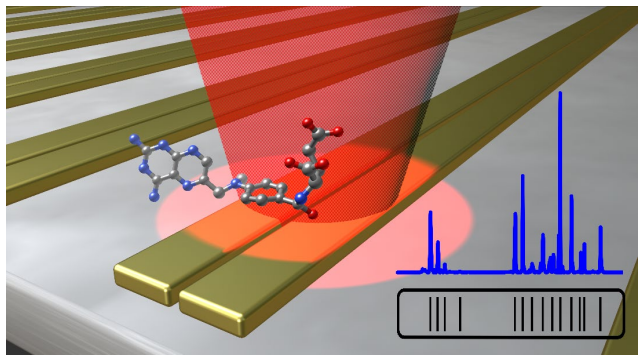

# Supplementary information

**Table S1:** Abbreviations used in the article.

| | |
|---|---|
| 3-MPA | 3-Mercaptopropionic acid |
| 4-ATP | 4-Aminothiophenol |
| BSA | **Bovine Serum albumin protein** |
| BT | Benzenethiol |
| DDW | double-distilled water |
| DMAB | 4,4'-dimercaptoazobenzene |
| DWGS | double wire grating substrate |
| EBSD | electron backscatter diffraction |
| EF | **E**nhancing factor |
| E-SERS | Electrically modulated SERS |
| FIB | focused ion-beam |
| LDOS | local density of states |
| M-Benzoate | Methyl benzoate |
| M-blue | Methylene blue |
| mc-Au | monocrystalline gold |
| pc-Au | polycrystalline gold |
| PL | Photoluminescence |
| PSO | Particle Swarm optimization |
| R6G | Rhodamine 6G |
| RS | Raman scattering |
| SAM | self-assembled monolayers |
| SEM | scanning electron microscopy |
| SERS | surface-enhanced Raman scattering |
| SWGS | single wire grating substrate |

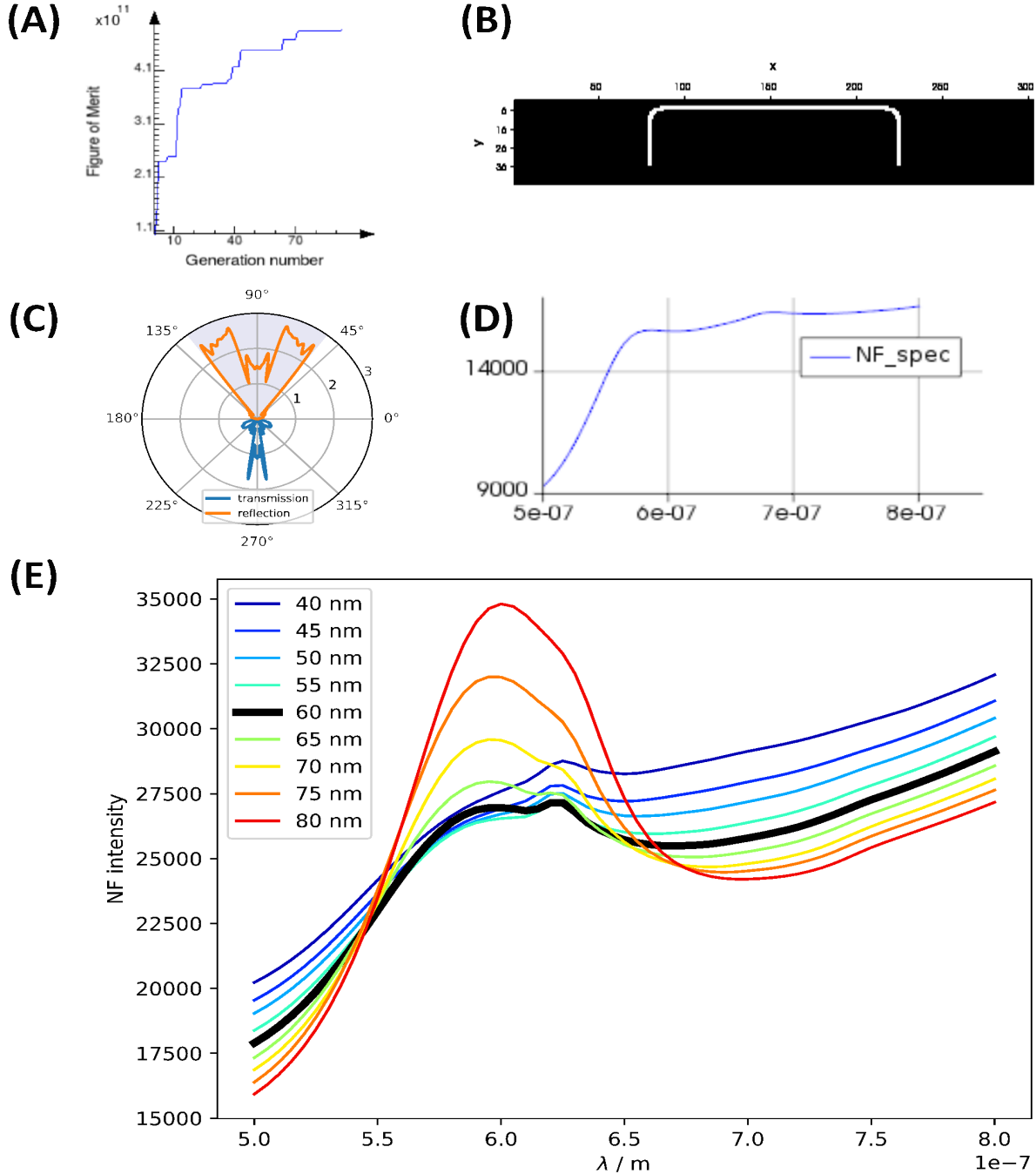

**Fig S1: Optimization of the DWGS geometry.** (A) Development of the grating fitness through 100 generations towards the geometry presented in the main paper. (B) Example of a mask for a single grating spanning two Yee-cells (= 4 nm) used to evaluate the near-fields only near the gold surface. (C) Far-field pattern for the DWGS at $\lambda = 635$ nm. It is clearly seen that the reflection is optimized to the numerical aperture of a NA = 0.6 objective (grey area). Due to reciprocity this also means maximal reception for the excitation. (D) Near-field spectrum for another optimization run, yielding the additional resonance at the emission wavelength. The overall intensity is lower than for the DWGS (compare with panel E). This result shows that several local fitness maxima can exist in any non-trivial optimization problem. (E) Near-field enhancement is shown for the optimized double grating dimensions, with increasing height from 40 to 60 nm. However, assessing the result is difficult as it also alters the far-field properties.

**Analytical enhancing factor calculation:**

The enhancing factor was determined by a dip-casting method of DWGS into a thiolation solution (experimental section) to perform a 4-ATP SAM followed by measuring the spectrum. For non-SERS, a reference spectrum was obtained by drop-casting of 15 mM 4-ATP onto a planar, non-grated gold surface. The spectra of both substrates were measured using the same settings. The enhancing factor was then calculated by averaging the 1077 cm$^{-1}$ integrated peaks of each substrate according to the formula:

$$EF = \frac{I_{SERS}}{I_R} \cdot \frac{N_{SERS}}{N_R}$$

Here, $I_{SERS}$ is the average integrated 1077 cm$^{-1}$ band retrieved using the DWGS substrate, and $I_R$ is the same integrated peak obtained on a planar gold surface under non-SERS conditions. The average number of molecules in the scattering volume for the plain Raman measurement is denoted as $N_R$. $N_{SERS}$ represents the coverage of the 4-ATP SAM on the grated SERS substrate, assumed 100% after 24 hr. [67] Since the drop casted reference sample was dried on the surface, the concentration of molecules per surface area was estimated to be 15 mM of 0.5 µl on approximately 1 mm$^2$, resulting in an estimated value of ~6±4×10$^{-7}$ mole/cm$^2$. Finally, $N_{SERS}$ is estimated to fall within the range of ~ 1×10$^{-9}$ mole/cm$^2$, as reported by other group.[67] The resulting SERS enhancing factor for the DWGS falls within the range of ~ 2.37±2.23×10$^6$ according to our protocols and sample preparation methods (experimental section). It is worth noting that previous investigations have estimated enhancement factors of around ~ 10$^6$ for these SAMs, which represents a level of SERS activity commonly observed in various experimental systems.[68]

**Raman 2D mapping:**

To investigate the relationship between morphology, plasmonic properties, and the spatial distribution of gold nanostructures in term of SERS-enhancing dimensions and hot-spots, we performed a scanning confocal 2D Raman mapping of 4-ATP SAM on of 5 × 5 µm$^2$ DWGS sample. The results obtained are presented in Fig. S2A-C. In the Raman mapping, each pixel in the matrix represents the integrated CCD counts of the total spectrum at 200 cm$^{-1}$-2500 cm$^{-1}$ (Fig. S1A), 4-ATP analyte 1077 cm$^{-1}$ peak (Fig. S2B) and 4-ATP 1570 cm$^{-1}$ peak (Fig. S2C). The mapping reveals predicted correlated hot-spots along the gold wires. Furthermore, periodic modulation due to the periodic milled gold wires, resulting in a periodic increase of approximately 3×10$^3$ cts in periodic manner as shown in Fig. S2D profile plot and SI short movie. Fig. S2 suggests that enhanced electromagnetic fields are localized around the wires and at junctions between adjacent wires. The average full width half maximum (FWHM) of the corresponding integrated peaks is approximately 337 nm ± 32 nm (Fig. S2D). The precise dimensions of two adjacent wires in the DWGS, including the small gap, amount to 290 nm, with each wire measuring 139/136 nm and the gap measuring 15 nm. The averaged calculated distance, obtained from Raman 2D mapping, between the center of the two enhancing wire-gap-wire nanostructures is approximately 840 ± 20 nm. This correlates with the actual distance from the center of one adjacent wire to the center of the next adjacent wire, which measures approximately 730 nm according to the SEM images (Fig. 2). Deviations between the measured values from the SEM images can be attributed to the optical limitations of the system. For example, large gap dimension between the wires (441 nm) is close to the theoretical lateral resolution limit of 430 nm according to Rayleigh criterion, which may explain the observed discrepancies in dimensions between the SEM data in Fig. 2 and the calculated distances in the 2D Raman mapping in Fig S2.

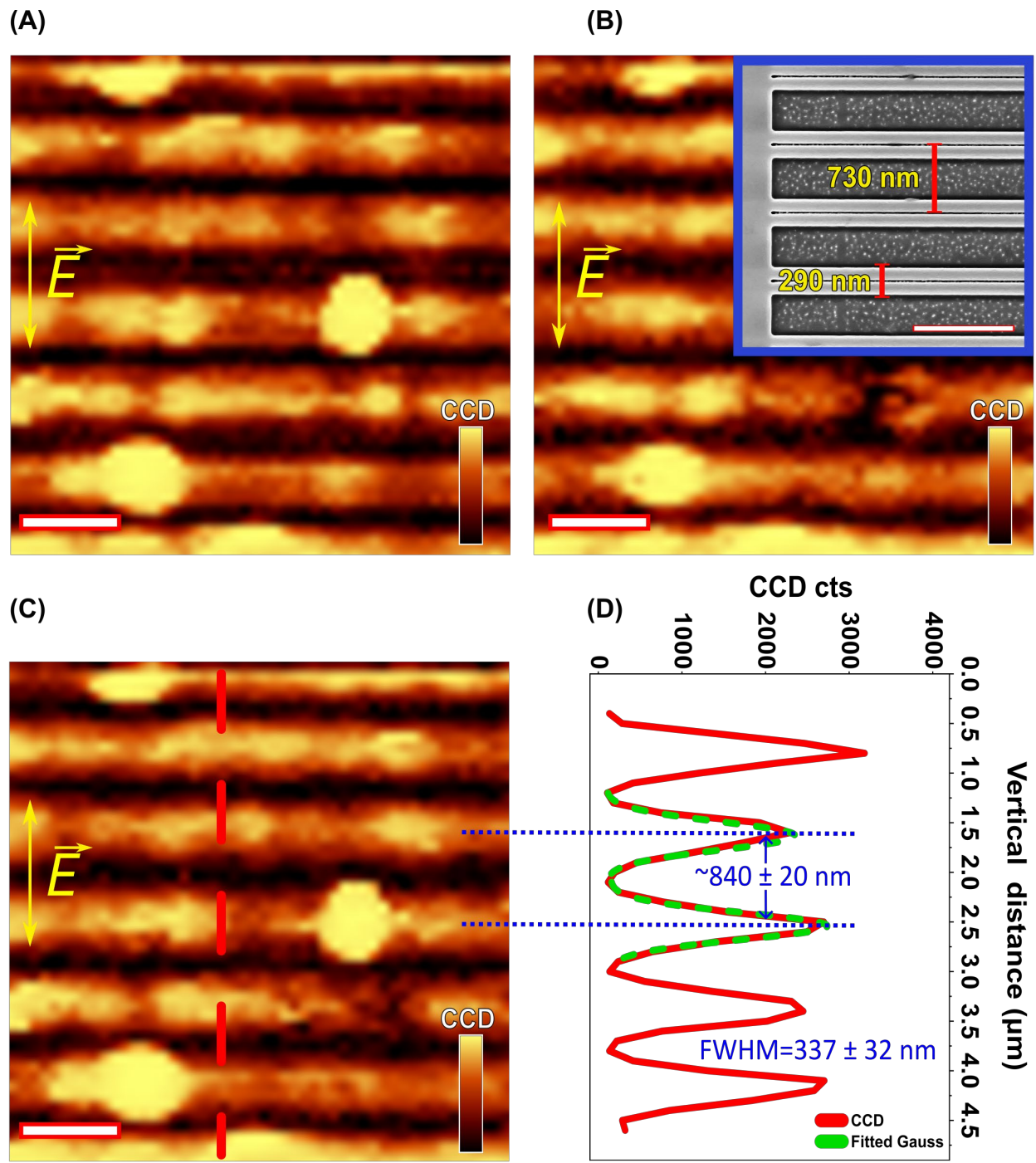

**Fig S2: 2D Raman mapping for 4-ATP-DWGS.** Panel (A)-(C) display a confocal 2D Raman mapping of the 4-ATP SAM chemically immobilized on the DWGS nanostructure. SEM inset in (B) upper part shows DWGS mapped geometry. The mapping matrix represents the integrated CCD counts of the 4-ATP spectrum within 200 cm$^{-1}$ - 2500 cm$^{-1}$ interval in (A), the integrated peak at 1077 cm$^{-1}$ in (B) and the integrated peak in (C). Panel (D) presents the quantitative changes in signal intensity of the 1570 cm$^{-1}$ band along the vertical red dashed line of (C). Scalebar 1 μm.

**2D Raman mapping experimental details.**

The mapping measurements were performed in WITec's application lab (WITec GmbH, Ulm, Germany) using a confocal microscope and Alpha300 R Raman system. A 633 nm excitation wavelength laser was utilized with an on-sample excitation power of 0.7 mW. A grating with 300 gr/mm was employed, and the measurements were conducted with a 100×/0.9 N.A. objective. The laser's linear polarization was directed perpendicular to gold wire to maximize the signal intensityThe substrate DWGS area scanned was 5 × 5 µm$^2$, resulting in a total of 50 × 50 pixels. Each pixel was acquired with steps of 100 nm and 1 second exposure time. The analysis of the acquired data was performed using the WITec Project and Project Plus system software (version 6.1.4.128). To generate a correlating 2D image, True Component Analysis algorithm was conducted, wherein the spectrum of interest (4-ATP) was selected from the detected components in the matrix. Subsequently, a fitting of the integrated spectral value of the 4-ATP spectrum was applied to each pixel. The resulting intensity profile plot on the right of Fig S2(D) represents the values along the red line X-axis (ranging from 0-4.9 µm) against the intensity of each pixel (with a step size of 100 nm). The gaussian curve fitting and mathematical analysis was done using the OriginPro 2023 software.

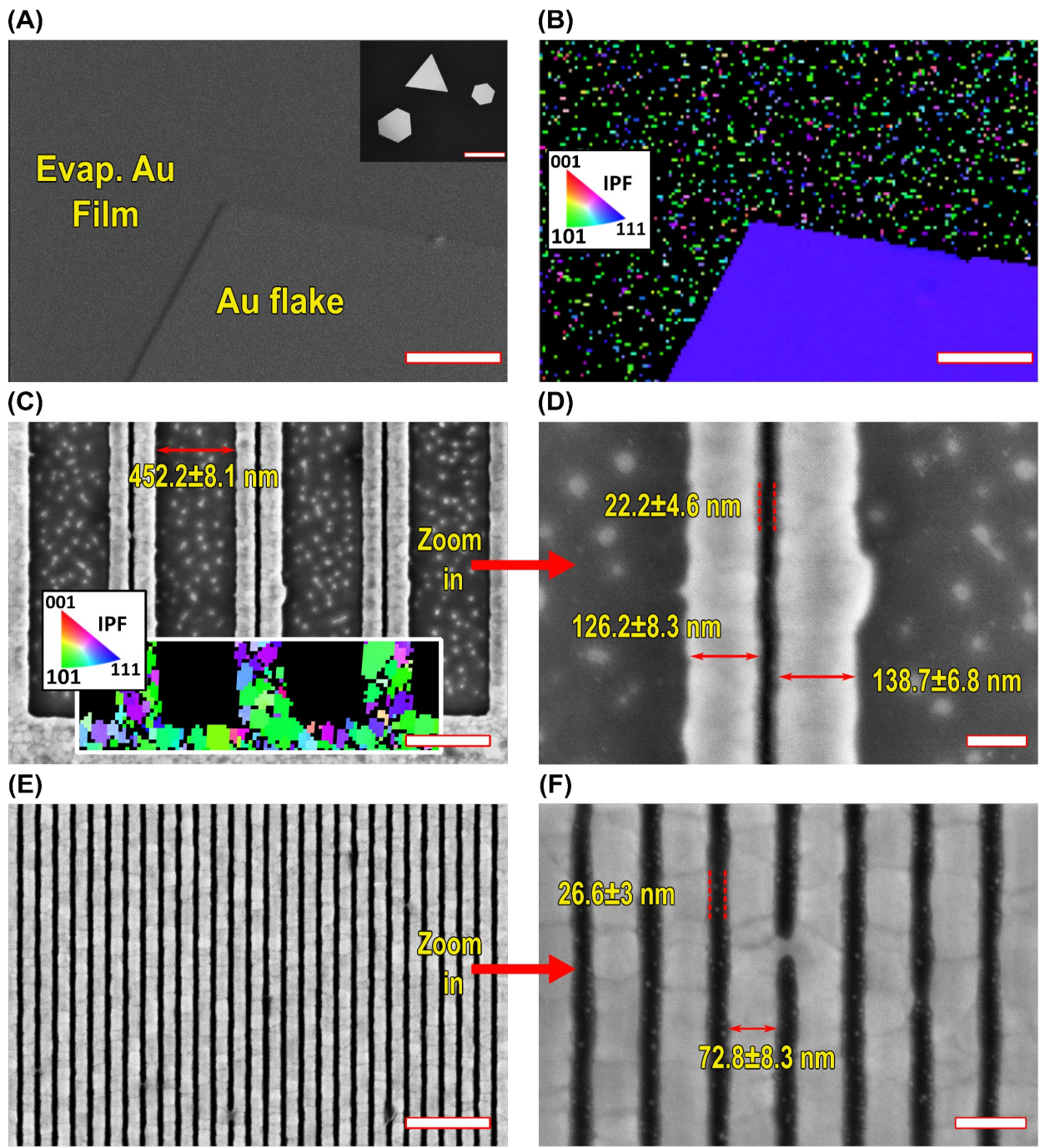

**Fig. S3: Au flake and evaporated Au SEM and EBSD characterization.** (A) SEM images of chemically synthesized monocrystalline Au flakes deposited on evaporated polycrystalline Au film. The upper rectangular shows a zoomed-out view of several gold flakes with different shapes. (B) Aligned IPF maps, calculated from the EBSD measurement, superimposed on the corresponding SEM image of (A), demonstrate that the chemically synthesized flakes are monocrystalline, while the evaporated gold film underneath is polycrystalline with various orientations and boundaries. The stereographic triangle of the IPF color map is located on the left side of figure (B). We present SEM images of evaporated gold SERS DWGS/SWGS fabricated using focused ion beam lithography. These images correspond to DWGS ((C), (D)) or SWGS ((E), (F)). In Figure (C), EBSD-aligned IPF maps are superimposed on the SEM image, highlighting the polycrystallinity by showing different crystal orientations. The glass appears black in the IPF map due to its amorphous structure. The scale bars for reference are as follows: (A) small upper rectangular - 30 μm, (A), (B) - 5 μm, (C), (E) - 500 nm, (D), (F) - 100 nm.

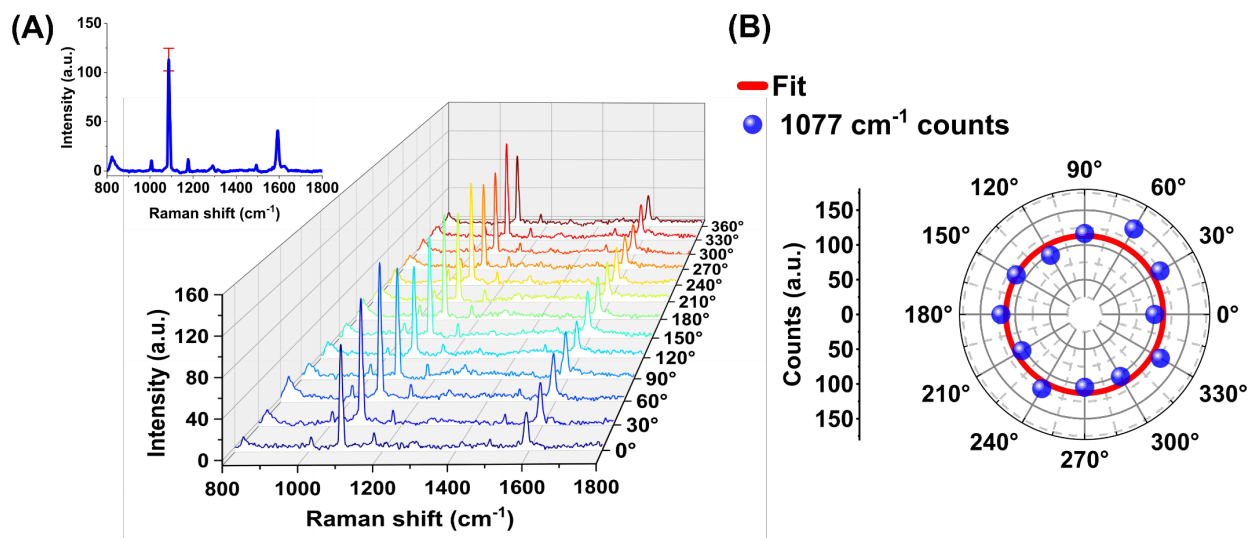

**Fig. S4: 4-ATP conventional RS against different polarization angels.** (A) The intensity of the distinct peak at 1077 cm$^{-1}$ of the analyte remained unchanged with variations in the electric field polarization angle. The average spectrum and relative standard deviation of 1077 cm$^{-1}$ peak are depicted on the upper left side of panel, with a relative standard deviation (RSD) of 11.4%. (B) The polar plot on the right shows the fitting of the 1077 cm$^{-1}$ Raman peak intensity as a function of the rotation angle. The blue dots represent the experimental data, and the red lines depict the theoretical fitting curve, highlighting the polarization independence of the conventional Raman scattering. Data were fitted by $y(\theta) = y_0$, where $y_0$ represent the average of 4-ATP Raman intensities in all angels.

### M-blue and 4-ATP spectra

Sensing low concentration of 4-ATP solution was demonstrated. The substrate was immersed in 4-ATP solution with concentration of 10$^{-12}$ M for 1 hour (experimental methods). Sub-monolayer coverage was insured, as 10$^{-12}$ M in 5 ml contains up to 5×10$^{-15}$ moles, and the gold substrate larger than 1 cm$^2$. Whereas 4-ATP SAM or small thiol molecules are in the range of ~ 1-27 × 10$^{-9}$ mole/cm$^2$.[67, 69, 70] Measured spectra is depicted in Fig. S5A. Different M-blue concentrations and accompanied calibration curve are depicted in Fig. S5B.

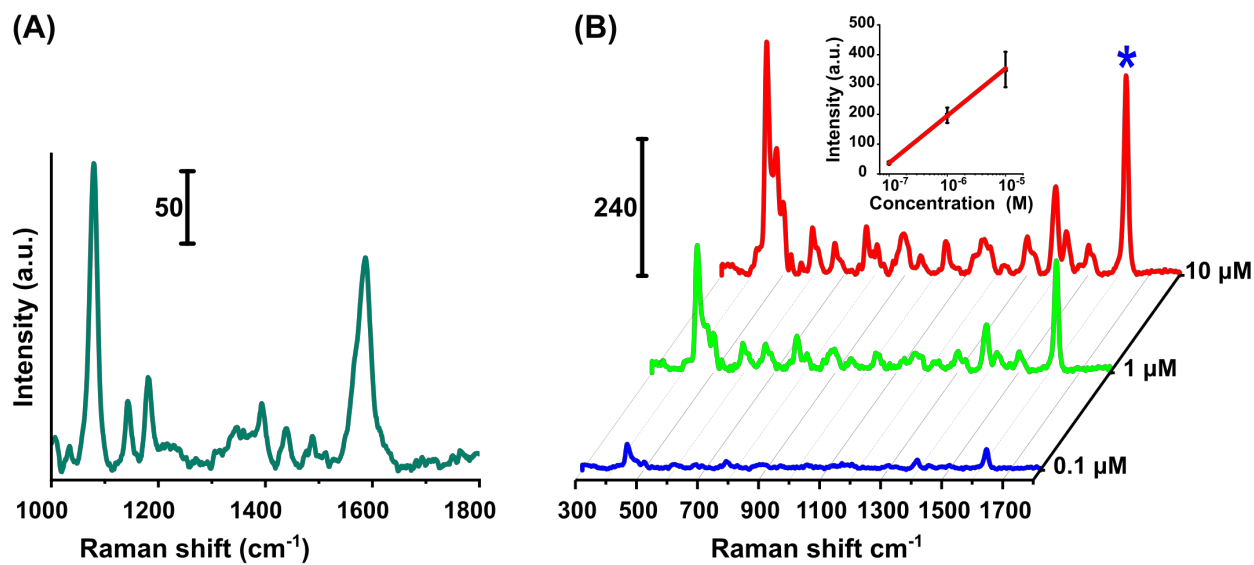

**Fig. S5: SERS of 4-ATP low concentration and M-blue different concentrations.** (A) 4-ATP spectrum obtained after dip-casting DWGS in picomolar concentration. Panel (B) The SERS intensities were measured at various M-Blue concentrations. The inset shows a linear calibration curve fitted within the range of 0.1 µm to 10 µm. The calibration curve represents the relationship between the M-Blue concentration (logarithmic) and the intensity of the M-Blue 1624 cm$^{-1}$ peak (indicated by the blue star).

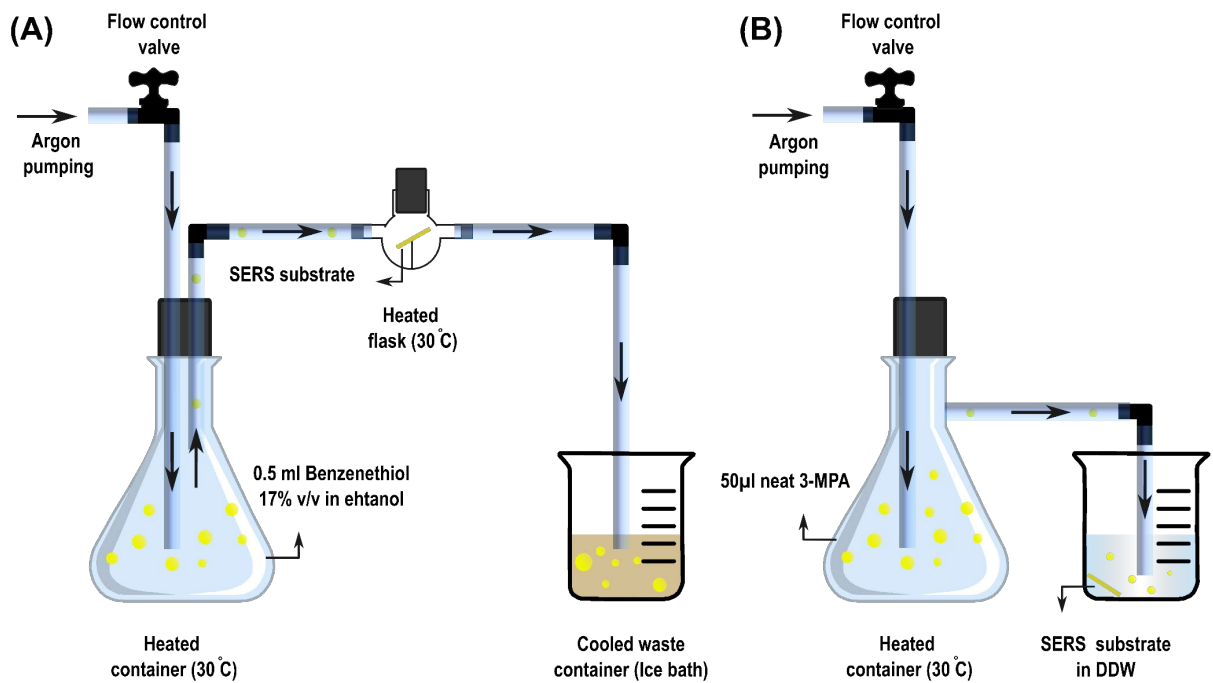

**Fig. S6: Schematics of gaseous volatiles sensing**. In (A), gas phase sensing procedure is depicted, illustrating the active flow of volatile gaseous Benzenethiol (BT) molecules through a glass tube directly to the SERS substrate heated flask (30 °C). The heating of the substrate flask prevents undesired condensation inside the flask/substrate. The substrate is positioned at a 30° tilt within the flask. In (B), the sensing of gaseous molecules is achieved by solubilizing volatile 3-MPA molecules in a chemical beaker containing a SERS substrate immersed in pure DDW. The substrate is positioned at an angle where the grating faces the solution. Heating for both systems is maintained within the range of 30 °C to enhance the volatility of the molecules. The boiling points of BT and 3-MPA are 169 °C and 110 °C, respectively.

**DWGS and planar gold substrates background signals**

Photoluminescent (PL) is known to be plasmon resonance-dependent and are considered as general background signals in typical SERS spectra.[71, 72] Here, we investigated the background signals of DWGS and planar non-grated gold substrates with and without the presence of the 4-ATP analyte. Our aim was to understand the shaping and influence of photoluminescence on the SERS signature of the analyte.

We obtained SERS spectra from both DWGS and planar non-grated substrates, while the corresponding photoluminescence spectra were acquired from the same substrates without the 4-ATP analyte. The results are depicted in Fig. S7 and Fig. S8.

The background signal from DWGS, perpendicular to the polarization, contributes to the SERS background signal of 4-ATP, as shown in Figure S8 (A) and (D) when exciting the gap plasmons in the grating through a dipolar resonance perpendicular to the wires. The photoluminescence is reduced when the grating is parallel to the polarization. The broad peak at 1600 cm$^{-1}$ is most likely the Raman of amorphous carbon,[73] which could be attributed to incomplete oxidation of organic contaminants either due to the cleaning technique or the laser.

All planar non-grated substrates, with or without the analyte, parallel or perpendicular to the polarization, exhibited similar spectra, noise, and photoluminescence, as shown in Fig. S8 (B).

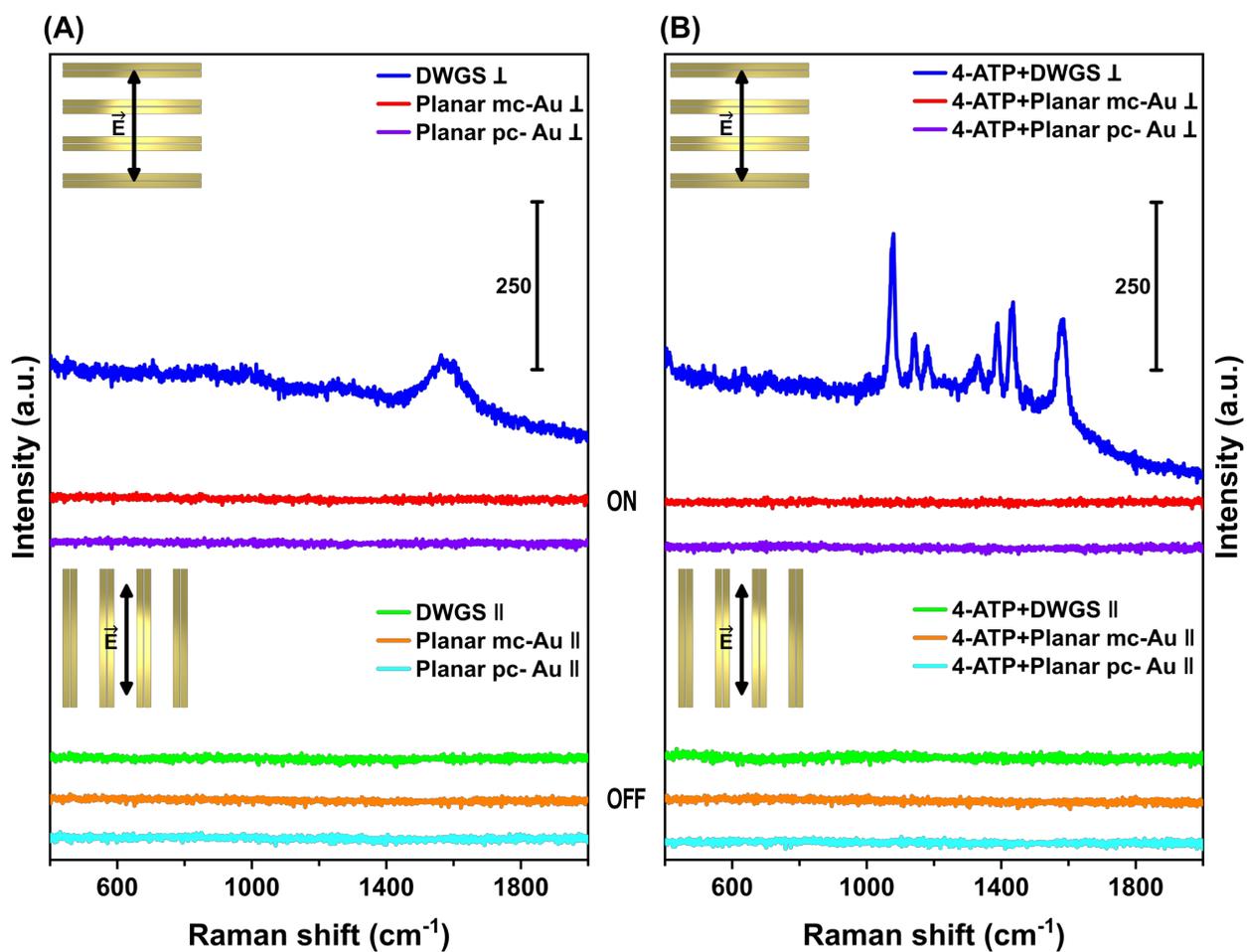

**Fig. S7: DWGS photoluminescence and background noise.** All spectra are presented without any spectral smoothing or baseline correction. Plots are shown with an offset for improved clarity. (A) Spectra of DWGS and planar non-grated gold substrate without the analyte, in both perpendicular (ON configuration) and parallel (OFF configuration) orientations to the polarization. (B) Spectra of the same substrates spectra after applying 4-ATP SAM.

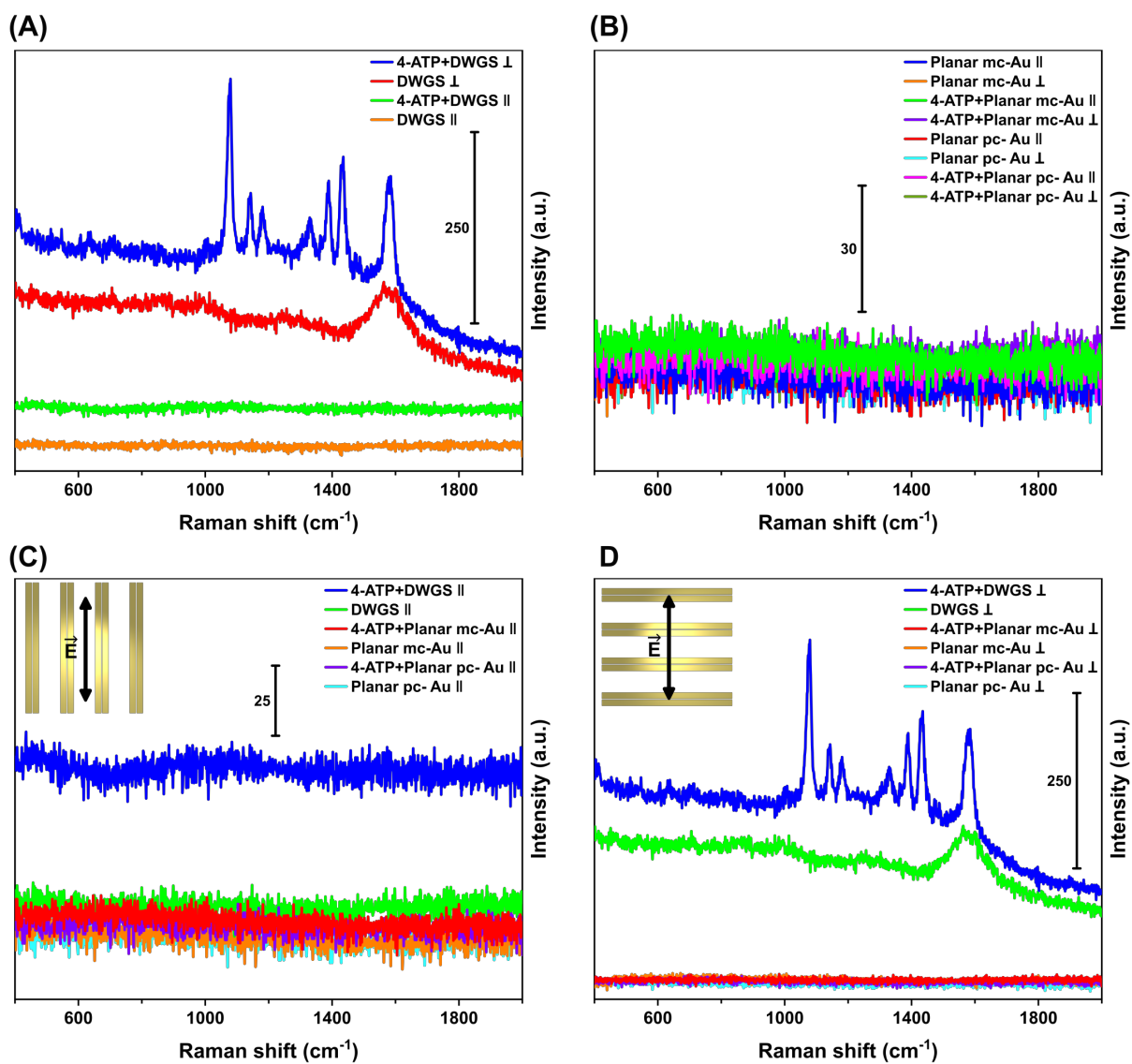

**Fig. S8: Comparison of DWGS photoluminescence and background noise.** Spectra are displayed on the same scale without offset for better comparison. (A) Spectra of all DWGS with and without the analyte. (B) Spectra of all planar gold substrates with and without the analyte. Comparison of spectra parallel to polarization and perpendicular to it are shown in (C) and (D), respectively.

34